\documentclass[manuscript]{aastex}

\usepackage[T1]{fontenc}
\usepackage[utf8]{inputenc}
\usepackage{lmodern}
\usepackage{graphicx}
\usepackage{amsmath}
\usepackage{textcomp}
\usepackage{changepage}
\usepackage{float}
\usepackage{amssymb}
\usepackage{rotating}
\usepackage{txfonts}
\usepackage{dblfloatfix}
\usepackage{multicol}
\usepackage{natbib}
\usepackage{color}
\usepackage[normalem]{ulem}

\makeatletter
\let\@dates\relax
\makeatother

\begin{document}

\title{The Role of Fission in Neutron Star Mergers and its Impact on the r-Process Peaks}

\author{M. Eichler\altaffilmark{1}, A. Arcones\altaffilmark{2,3}, A. Kelic\altaffilmark{3}, O. Korobkin\altaffilmark{4}, K. Langanke\altaffilmark{2,3}, T. Marketin\altaffilmark{5},\\
G. Martinez-Pinedo\altaffilmark{2,3}, I. Panov\altaffilmark{6,1}, T. Rauscher\altaffilmark{7,1}, S. Rosswog\altaffilmark{4}, C. Winteler\altaffilmark{8}, N.T. Zinner\altaffilmark{9},\\
F.-K. Thielemann\altaffilmark{1}}
\altaffiltext{1}{Department of Physics, University of Basel, Klingelbergstrasse 82, 4056 Basel, Switzerland}
\email{marius.eichler@unibas.ch}

\altaffiltext{2}{Institut f\"ur Kernphysik, Technische Universit\"at Darmstadt, Schlossgartenstrasse 2, D-64289 Darmstadt, Germany}
\altaffiltext{3}{GSI Helmholtzzentrum f\"ur Schwerionenforschung GmbH, Planckstrasse 1, D-64291 Darmstadt, Germany}
\altaffiltext{4}{The Oskar Klein Centre, Department of Astronomy, AlbaNova, Stockholm University, SE-106 91 Stockholm, Sweden}
\altaffiltext{5}{Department of Physics, Faculty of Science, University of Zagreb, 10000 Zagreb, Croatia}
\altaffiltext{6}{SSC RF ITEP of NRC "Kurchatov Institute", Bol'shaya Cheremushkinskaya 25, 117218 Moscow, Russia}
\altaffiltext{7}{Centre for Astrophysics Research, School of Physics, Astronomy and Mathematics, University of Hertfordshire, Hatfield AL10 9AB, UK}
\altaffiltext{8}{Institut Energie am Bau, Fachhochschule Nordwestschweiz, St. Jakobs-Strasse 84, 4132 Muttenz, Switzerland}
\altaffiltext{9}{Department of Physics and Astronomy, Aarhus University, Ny Munkegade, bygn. 1520, DK-8000 Aarhus C, Denmark}

\begin{abstract}
Comparing observational abundance features with nucleosynthesis predictions of stellar evolution or explosion simulations can scrutinize two aspects:
(a) the conditions in the astrophysical production site and (b) the quality of the nuclear physics input utilized. We test the abundance features of r-process nucleosynthesis 
calculations for 
the dynamical ejecta of neutron star merger simulations based on three different nuclear mass models: The Finite Range Droplet Model (FRDM), the 
(quenched version of the) Extended Thomas Fermi Model with Strutinsky Integral (ETFSI-Q), and the Hartree-Fock-Bogoliubov (HFB) mass model.
We make use of corresponding fission barrier heights and compare the impact of four different fission fragment distribution models on the final r-process abundance distribution. In particular, we explore the abundance distribution in the second r-process peak and the rare-earth sub-peak as a function of mass models and fission fragment distributions, as well as the origin of a shift in the third r-process peak position. The latter has been noticed in a number of merger nucleosynthesis predictions. We show that the shift occurs during the 
r-process freeze-out when neutron captures and $\beta$-decays compete and an (n,$\gamma$)-($\gamma$,n) equilibrium is not maintained anymore. 
During this phase neutrons originate mainly from fission of material above $A = 240$. We also investigate the role of $\beta$-decay half-lives from recent theoretical advances, which lead either to a smaller amount of fissioning nuclei during freeze-out or a faster (and thus earlier) release of fission neutrons, which can (partially) prevent this shift and has an impact on the second and rare-earth peak as well.

\end{abstract}


\section{Introduction}
The rapid neutron capture process (\textit{r-process}) is responsible for the production of about half of the elements heavier than iron in our universe. It is characterized by fast neutron captures in comparison to $\beta$-decays and follows a path in the nuclear chart that runs close to the neutron drip line. Its basic mechanism was already suggested by \citet{bbfh1957} and \citet{cameron1957}.

As of today, the astrophysical site(s) of the r-process remain(s) uncertain, but metal-poor stars with enriched r-process material offer valuable clues about the nature of the r-process source(s). Observations reveal that the [Eu/Fe]\footnote{[X/Fe] = $\log\left(Y_X/Y_{Fe}\right)_{Star}-\log\left(Y_X/Y_{Fe}\right)_{\odot}$} ratios of the oldest stars are scattered over several orders of magnitude, 
while the scatter decreases for younger and correspondingly less metal-poor stars \citep{cowan2004}. Europium is exclusively produced by the r-process and is therefore used as an indicator of r-process material enrichment.
On top of that, the overall element abundance pattern of heavy (``strong'') r-process nuclei follows the solar one with remarkable accuracy \citep{sneden2009}. In combination with the large scatter in [Eu/Fe] for low-metallicity stars, this points to a rare event, responsible for the production of heavy r-process material \citep{sneden2008,roederer2012}.
On the other hand, various intermediate-mass r-process elements up to Europium are observed in almost all stars, albeit at lower levels \citep{honda2007,roederer2013}.
This argues for an additional frequent event which can account for such a ``weak'' r-process signature. Regular core-collapse supernovae may well 
be the origin, as the neutrino wind could generate a slightly neutron-rich environment \citep{roberts2012,martinez2012}, but 
does not provide the entropies required for the operation of a strong r-process in slightly neutron-rich conditions.

A very rare and special class of supernovae is most likely powered by the so-called magnetorotational explosion mechanism \citep{bisnovatyi-kogan1970, leblanc1970, meier1976, burrows2007}. This mechanism in particular produces polar outflows with neutron-rich conditions ($Y_e \approx 0.2$), favorable for an r-process \citep{nishimura2006,fujimoto2008,winteler2012,nishimura2015}, and could be responsible for the r-process in the early evolution of galaxies.

In parallel, compact object mergers have long been suspected to be an alternative site for r-process nucleosynthesis \citep{lattimer1974,meyer1989,freiburghaus1999,roberts2011,goriely2011,korobkin2012,bauswein2013,rosswog2014}.
For most recent results see also \cite{just2014}, \cite{mendoza2014}, and \cite{wanajo2014}.
The combination of very low-$Y_e$ material and rapid expansion of the ejecta guarantees the occurrence of a strong r-process. 
Several studies \citep{goriely2011,korobkin2012,bauswein2013} uncovered remarkable astrophysical robustness of the abundance yields produced in the dynamical ejecta of neutron star mergers (NSM) and mergers of a neutron star with a black hole for a given nuclear input. This insensitivity of the abundance pattern to the parameters of the merging system is explained by an extremely low-$Y_e$ environment, which guarantees the occurrence of several fission cycles before the r-process freezes out. However, recent NSM simulations that also account for the neutrino-driven wind and/or viscous disk ejecta at a later stage of the merger find a much broader range of $Y_e$-values for the ejecta \citep{rosswog2014,just2014,perego2014,wanajo2014}.

In this study, we revisit the nucleosynthesis in the dynamical ejecta of NSMs of \cite{korobkin2012}, by using the ETFSI-Q \citep{aboussir1995,pearson1996} and HFB-14 \citep{goriely2008,goriely2009} mass models in addition to the FRDM mass model \citep{moeller1995}.
Our main focus is the effect of fission on the r-process path through the very neutron-rich, unstable nuclei, utilizing corresponding fission barriers (\citealt{myers1999} for FRDM and \citealt{mamdouh1998} for ETFSI-Q and HFB, as discussed in \citealt{panov2010}) and four fission fragment distribution predictions.
We also follow the decay back to stability during the r-process freeze-out and the competition between late neutron captures and neutron release by fission and $\beta$-decays of heavy nuclei (see similar discussions of freeze-out effects without including fission in \citealt{mumpower2012} and references therein).
Late neutron captures have a direct effect on the final position of the third r-process peak 
(also seen in \citealt{goriely2013} and \citealt{goriely2015}). To study the dependence of the final abundance distribution on the freeze-out characteristics, we pick a typical trajectory from the same database of trajectories\footnote{http://compact-merger.astro.su.se/} that was used in \citet{korobkin2012}.
We include the following three fission modes: spontaneous, $\beta$-delayed and neutron-induced fission, as described in detail in \citet{panov2008,panov2010} and \citet{petermann2012}.

Fission has become fundamental to understand the r-process in compact binary mergers. However, the study of fission fragments has been ongoing long before. Soon after the discovery of the neutron and proton shell 
structure and the development of the spherical-shell model, the mass distribution of fission 
fragments was linked to shell closures in the daughter nuclei \citep{fong1956}. The first quantitative predictions for fission fragment 
distributions were done using a statistical scission-point model (\citealt{wilkins1976}; see \citealt{steinberg1978} for the impact on r-process nucleosynthesis).
Recent advances are discussed in \cite{tatsuda2007}, \cite{panov2008}, \cite{kelic2008}, \cite{goriely2013}, and \cite{goriely2015}, with differences for the predicted mass distributions revealing the remaining uncertainties in present fission calculations.

In the present paper we explain in section 2 the method and parameters used in our nucleosynthesis calculations. The detailed results are 
presented in section 3 and summarized in section 4, giving also an outlook for the need of future studies.

\section{Nucleosynthesis Calculations}
\label{method}
\subsection{Basic Input and Conditions in Ejecta Trajectories}
We utilize the extended nuclear network \sc Winnet \rm \citep{winteler2012} with more than 6000 isotopes up to Rg. Our sets of reaction rates utilized are 
based on masses from the \textit{Finite Range Droplet Model} (FRDM; \citealt{moeller1995}), the \textit{Extended Thomas Fermi Model
with Strutinsky Integral} (with shell quenching) (ETFSI-Q; \citealt{aboussir1995,pearson1996}), both in combination with the statistical model calculations of \citet{rauscher2000} for $Z\leq 83$, and on the \textit{Hartree-Fock-Bogoliubov Model} (HFB-14; \citealt{goriely2008,goriely2009}), respectively.
Theoretical $\beta$-decay rates are taken from \cite{moeller2003}, experimental data from the nuclear database \cite{nudat2}. The neutron capture rates on heavy nuclei ($Z>83$) as well as the neutron-induced fission rates are from \citet{panov2010}, while the $\beta$-delayed fission rates are taken from \citet{panov2005}. In our calculations, we refer to the combined application of these basic sets of reaction rates as \textit{original}. We also test the effect of very recent advances in $\beta$-decay half-life predictions by \cite{marketin2015} and \cite{panov2015}.

We have performed r-process calculations for a neutron star merger scenario with two $1.4$~M$_{\odot}$ neutron stars \citep{rosswog2013,korobkin2012}. We use 30 representative fluid trajectories, covering all the conditions in the ejected matter and providing the temperature, density and electron fraction within the ejected material up to a time of $t_0 = 13$ ms. We start our nucleosynthesis calculations after the ejecta have begun to expand and the temperature has dropped to 10~GK. For $t > t_0$ we extrapolate, using free uniform expansion for radius, density and temperature:
\begin{align}
r(t) &= r_0 + t v_0 \\
\rho(t) &= \rho_0 \left( \frac{t}{t_0} \right)^{-3} \\
T(t) &= T[S,\rho(t),Y_e(t)],
\end{align}
with radius $r$, time $t$, velocity $v$, density $\rho$, temperature $T$, entropy $S$ and electron fraction $Y_e$ of the fluid element. The index $0$ denotes the values at $t_0$. The temperature is calculated at each timestep, using the equation of state of \citet{timmes2000}.

Our network accounts for heating due to nuclear reactions \citep{freiburghaus1999}, using
\begin{equation}
kT\frac{dS}{dt} = \epsilon_{th} \dot{q}
\end{equation}
to calculate the entropy increase caused by thermal heating, where $\dot{q}$ is the energy generated due to nuclear reactions. We choose a heating 
efficiency parameter of $\epsilon_{th} = 0.5$ (introduced in \citealt{metzger2010a}), corresponding to about half of the $\beta$-decay energy 
being lost via neutrino emission. The efficiency for neutron captures and fission processes should be higher, as none of the released energy escapes. However, the energy release in neutron captures is small due to small neutron-capture Q-values along the r-process path (compared to large $\beta$-decay Q-values), and the
abundances of heavy fissioning nuclei are small in comparison to the majority of nuclei in the r-process path. Thus, while the heating via beta-decays dominates, the exact value of 
$\epsilon_{th}$ is difficult to determine. In the case of extremely neutron-rich dynamic NSM ejecta, the final abundances are, however, quite 
insensitive to its value \citep{korobkin2012}.

\subsection{The Treatment of Fission in Nuclear Networks}
The fission fragment distribution depends on the nuclear structure of the fissioning nucleus as well as that of the fission products, e.g., the shell structure of nuclei far from stability. The fission products can be 
predicted statistically by assigning a probability to each possible fission channel. Rates for the various fission channels considered and the associated yield distributions are crucial for r-process studies in NSMs. 
In each fission reaction there is a possibility of several fission neutrons to be emitted. While the number of fission neutrons has been measured to be $2-4$ for experimentally studied nuclei, it is known to increase 
with mass number as heavy nuclei become more neutron-rich \citep{steinberg1978}. Additional neutrons can be emitted as the fission fragments decay towards the r-process path \citep{martinez-pinedo2006}.

For our NSM nucleosynthesis calculations we employ four different fission fragment distribution models: (a) \citet{kodama1975}, 
(b) \citet{panov2001}, (c) \citet{panov2008}, and (d) ABLA07 \citep{kelic2008,kelic2009}. The first one is a relatively simple parametrization that does not 
take into account the release of neutrons during the fission process. The second and third are parametrizations guided by experimental data. The 
number of released neutrons in \cite{panov2008} has been estimated as a function of charge and mass number of the fissioning nucleus and can reach 10 per fragment for 
nuclei near the drip line. The ABLA07 model is based on a statistical model considering shell effects in the fragments from theoretical predictions and has been tested to provide an accurate description 
of known fission data including the number of released neutrons \citep{benlliure1998,gaimard1991,kelic2009}. It also includes the reproduction of fragment distributions from extended heavy ion collision yields \citep{kelic2008}, and therefore goes much beyond the areas in the nuclear 
chart where spontaneous, beta-delayed or neutron-induced fission yields are known experimentally. 
Further applications of recent fragment models (GEF and SPY) are also employed in \cite{goriely2015}. However, these models have either been restricted so far to not very neutron-rich nuclei ($A/Z < 2.8$ and $N < 170$; GEF), or have not yet been the subject of the same severe tests as ABLA07 (both GEF and SPY).

Based on NSM simulations of \cite{rosswog2013}, we present a comparison of the abundance features resulting from utilizing the different fission fragment models in section~\ref{results} (see also \citealt{eichler2013}).

\subsection{Beta-Decay Rates of the Heaviest Nuclei and their Relation to the Release of Fission Neutrons}
In the following section(s), containing results of our nucleosynthesis calculations, it will become apparent that not only the choice of a mass model and the treatment of fission is important for the final
abundances. It also turns out that the neutrons released during fission and from highly neutron-rich fission products play an important role for the late r-process freeze-out and the final abundance distribution. One of the main features is related
to the timing between neutron release in fission and from decay products and the final neutron captures, possibly occurring after the freeze-out from (n,$\gamma$)-($\gamma$,n) equilibrium. The role of beta-decays (especially of the heaviest
nuclei) is essential for the speed to produce heavy (fissioning) nuclei, as well as the time duration for which fissioning nuclei still exist.
There are experimental indications \citep{domingo2013,morales2014,kurtukian2014} that the half-lives of nuclei around $N=126$ are shorter than predicted by the FRDM+QRPA approach \citep{moeller2003}. Several theoretical calculations based on an improved treatment of forbidden transitions suggest a similar behavior \citep{zhi2013,suzuki2012}. \cite{caballero2014} have explored the impact on r-process calculations utilizing the half-live predictions of \cite{borzov2011}. Faster $\beta$-decay rates for nuclei in the mass region $Z>80$ are also supported by the recent theoretical work of \cite{panov2015} and \cite{marketin2015}.

\section{Results}
\label{results}
This section focuses on several aspects entering r-process nucleosynthesis in NSM ejecta: a comparison of different fission fragment distribution and mass models and their impact on the second r-process peak (section~\ref{ffd}), a discussion of the late capture of fission neutrons and the impact on the position of the r-process peaks (section~\ref{thirdpeak}), and the overall combined effect of mass models, fission, and $\beta$-decay half-lives on the abundance distribution for $A >120$ , i.e., including the second and the third peak (section~\ref{global}).
\subsection{The Effect of Fission Fragment Distributions}
\label{ffd}
\begin{figure*}[t!]
  \includegraphics[width=0.5\textwidth]{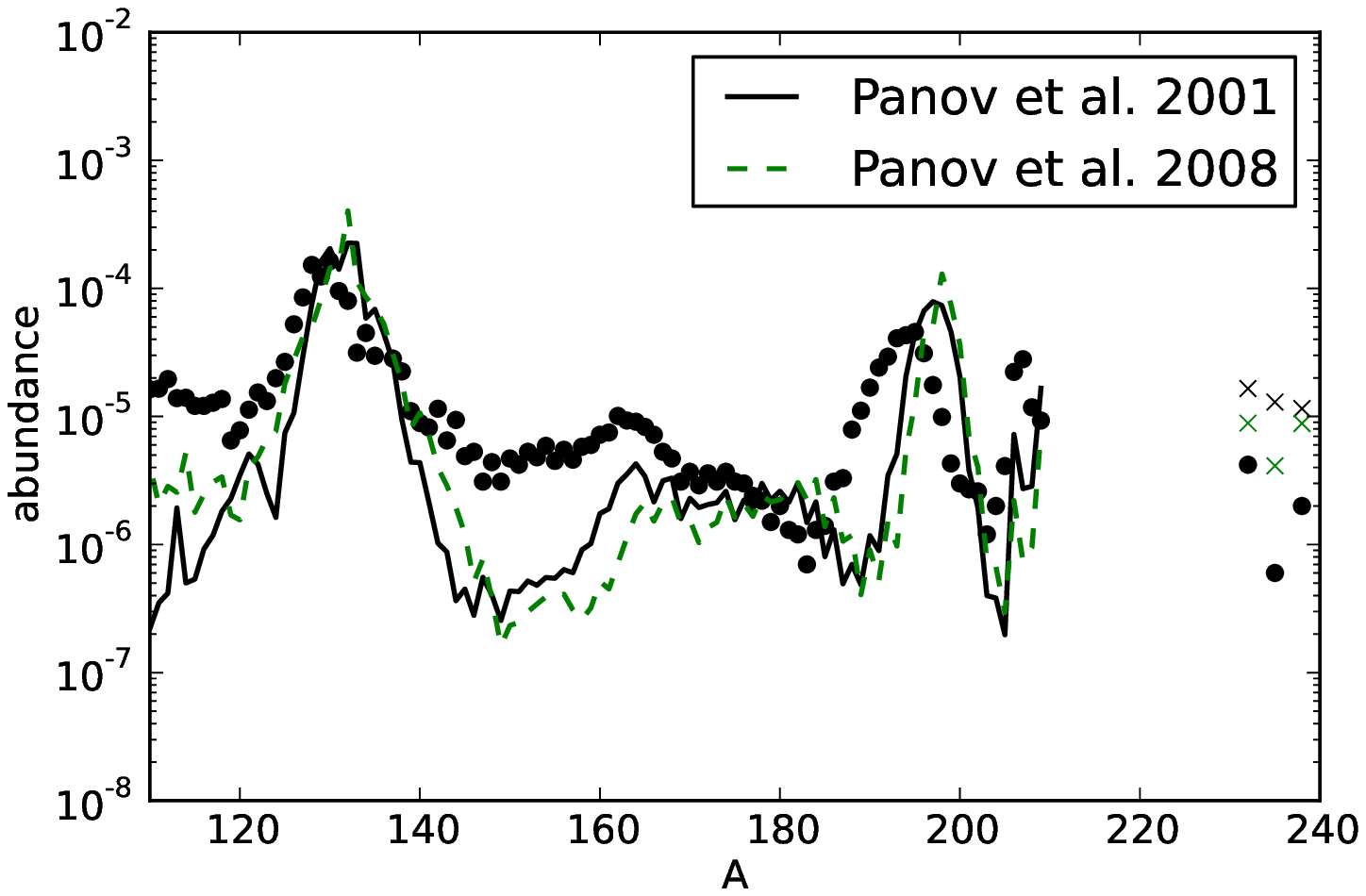}
  \label{ffdm13}
  \includegraphics[width=0.5\textwidth]{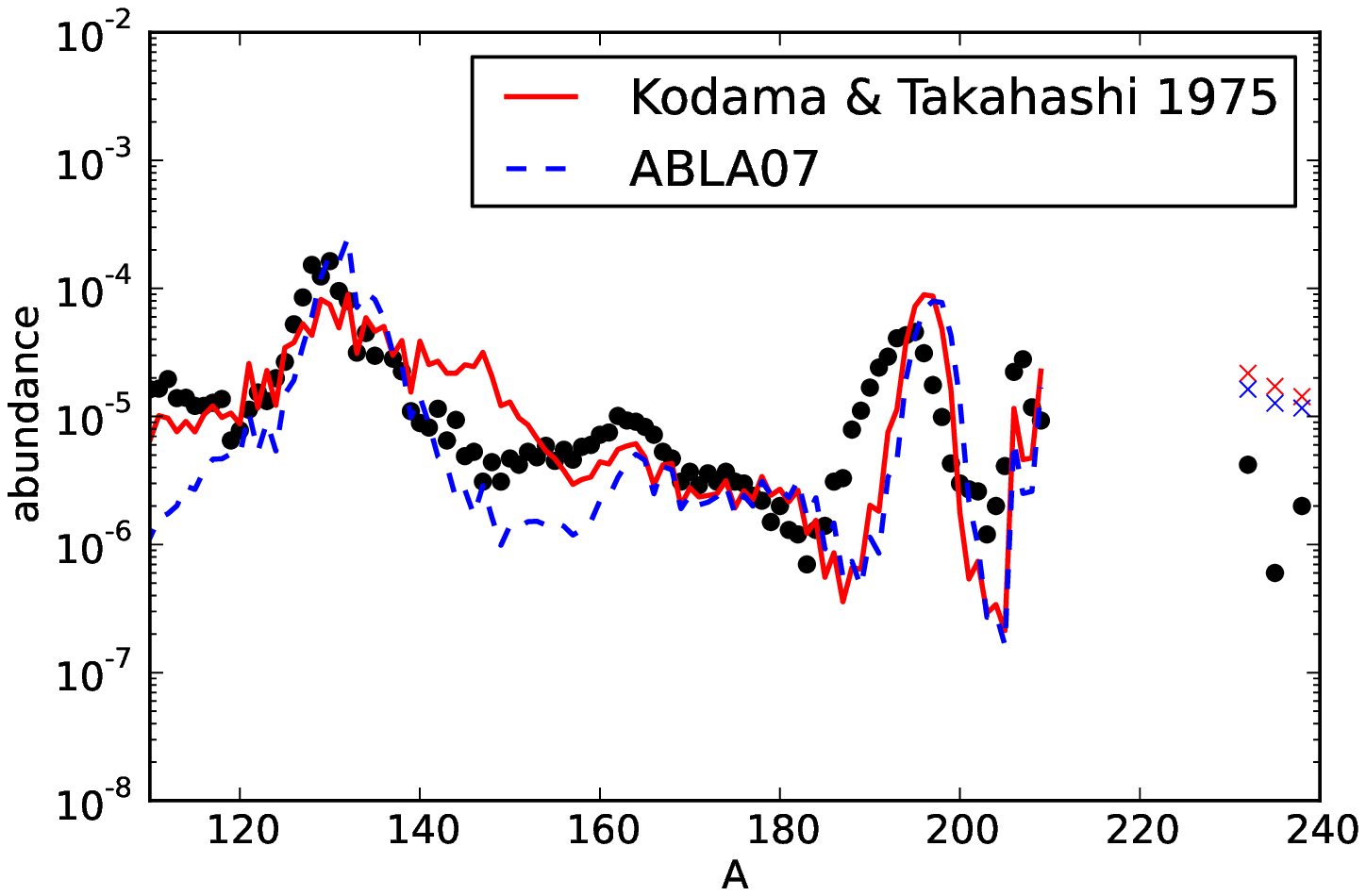}
  \label{ffdm24}
  \caption{Final abundances of the integrated ejecta around the second and third peak for a NSM \citep{rosswog2013,korobkin2012} at a simulation time $t=10^6$~s, employing the FRDM mass model combined with four different fission fragment distribution models (see text). For reasons of clarity the results are presented in two graphs. The abundances for Th and U are indicated by crosses. In the left-hand panel the lower 
crosses belong to the Panov et al. (2008) model (dashed line), while the lower crosses in the right-hand panel belong to the ABLA07 distribution model (dashed line). The dots represent the solar r-process abundance 
pattern \citep{sneden2008}.}
\label{ffdm}
\end{figure*}

The effect of adopting different fission fragment distribution models, all in combination with the FRDM mass model for the r-process calculations, is illustrated in Fig.~\ref{ffdm}. It shows final abundances of the NSM ejecta in the atomic mass range $A= 110 - 210$. Two of the fragment distributions \citep{panov2001,kodama1975} have already been used 
for NSM calculations before \citep{korobkin2012} (see also \citealt{bauswein2013}, utilizing the latter of the two, but with fragment mass and charge asymmetry derived from the HFB-14 predictions; see \citealt{goriely2011}), while the other two \citep{panov2008,kelic2008} have been newly implemented for the present 
calculations. All our results are compared to the solar r-process abundance pattern (\citealt{sneden2008}).

\begin{figure*}[p!]
%  \centering
  \includegraphics[width=0.55\textwidth]{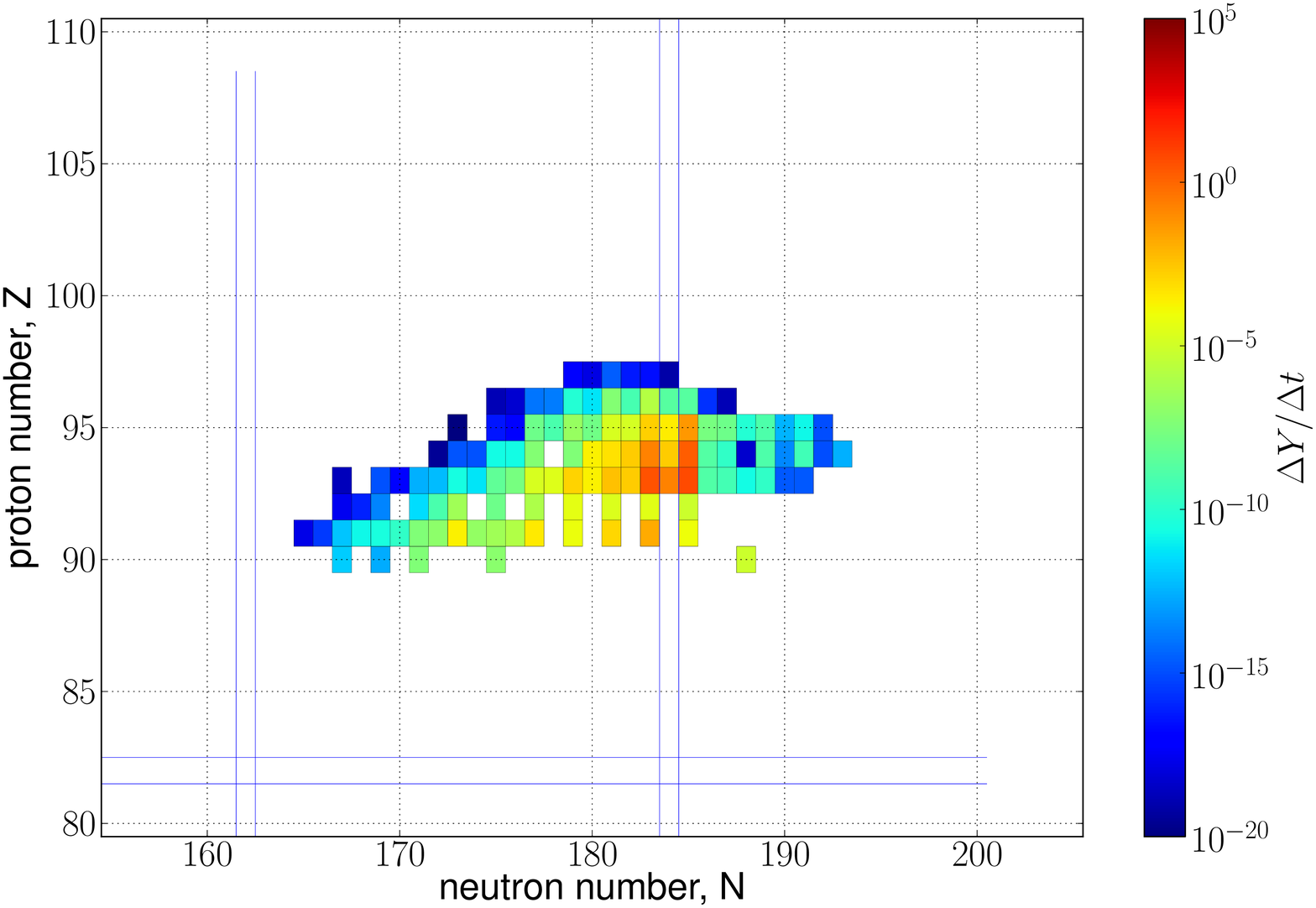}
  \label{beta_fission}
  \includegraphics[width=0.55\textwidth]{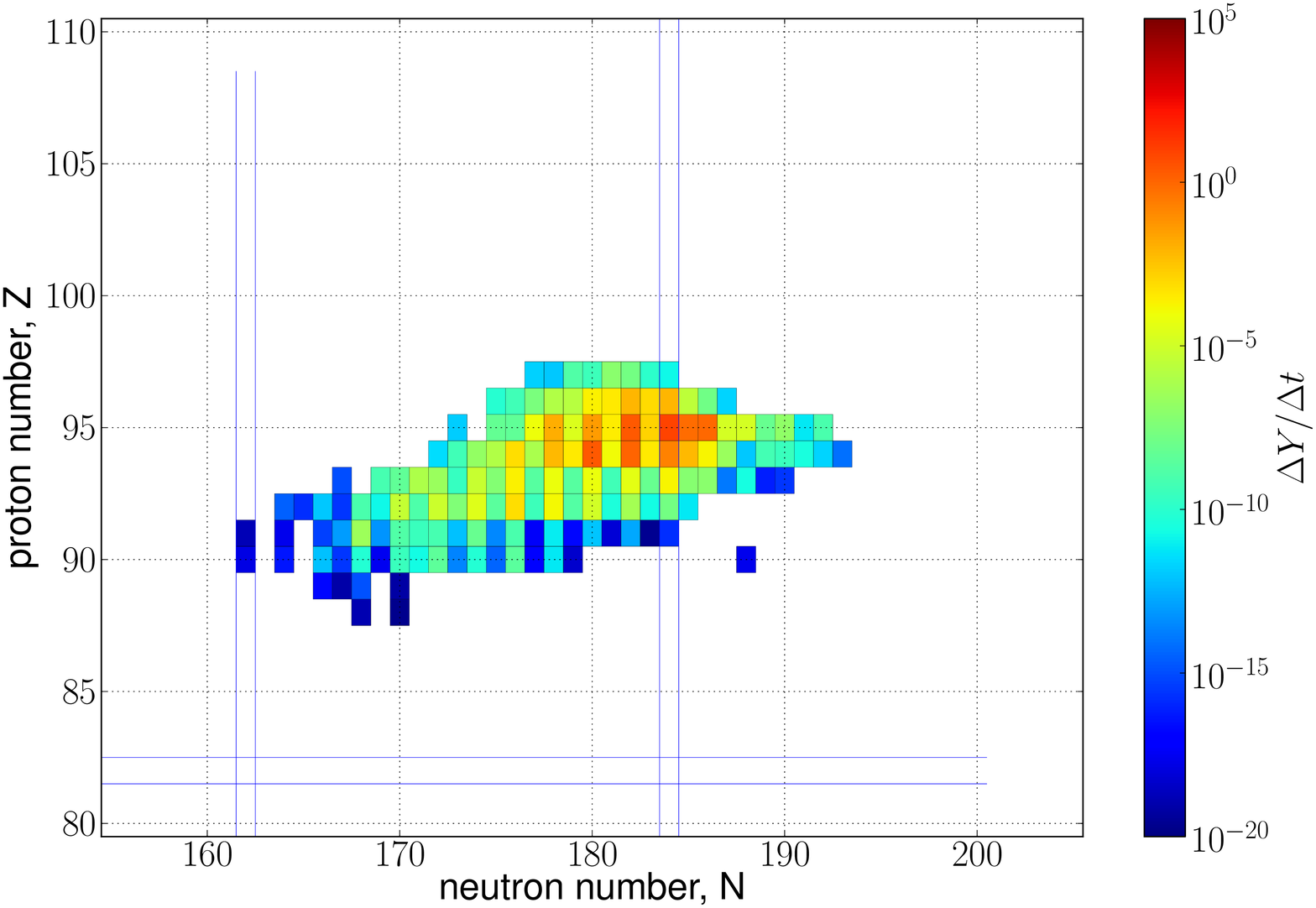}\\
  \label{ninduced_fission}
  \hspace{1.5cm}
  \includegraphics[width=0.8\textwidth]{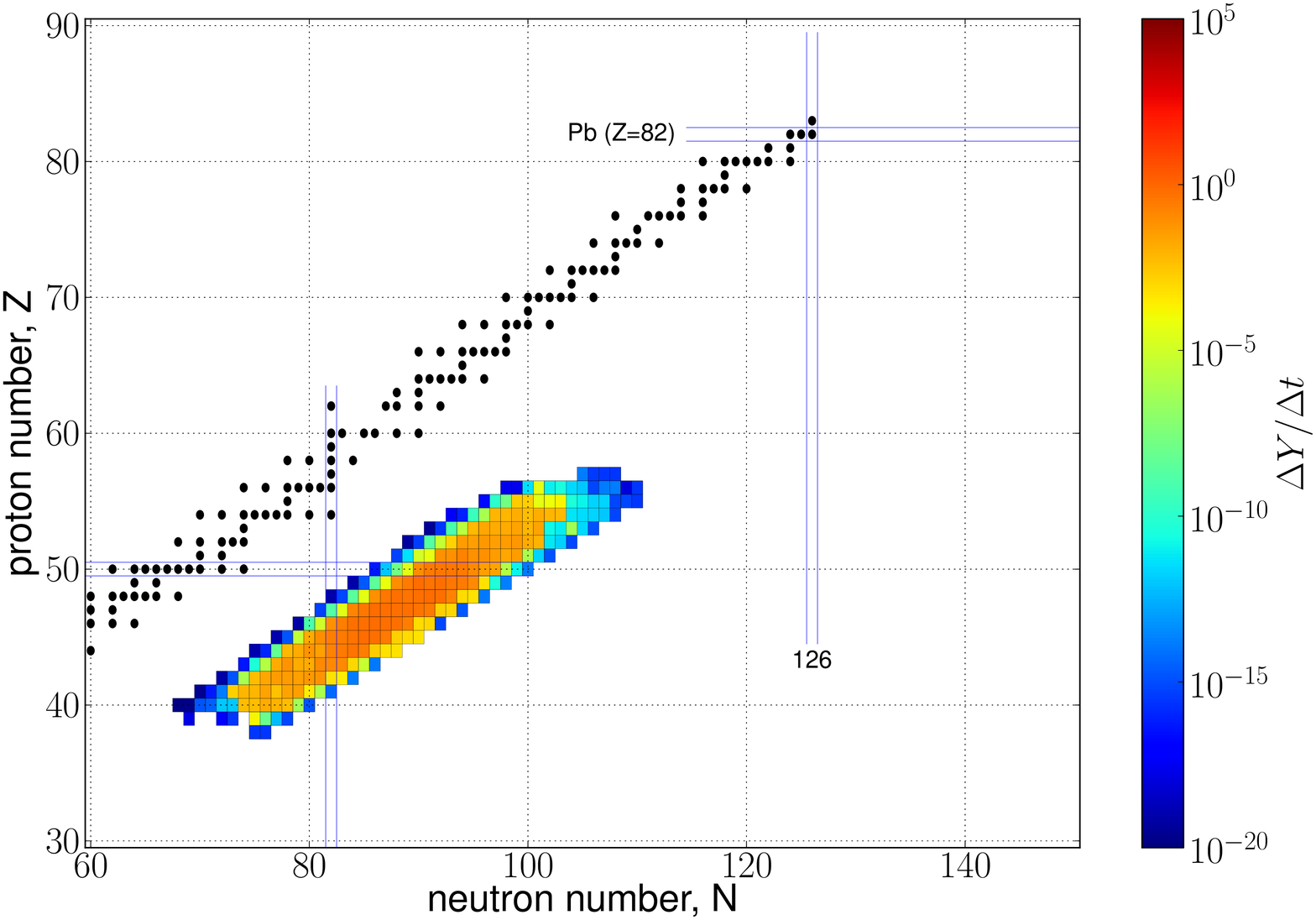}
  \label{fragments_abla}
  \caption{Fission rates (at $t = 1$s) in s$^{-1}$ for $\beta$-delayed (top left) and neutron-induced fission (top right) at freeze-out from (n,$\gamma$)-($\gamma$,n) equilibrium for one representative trajectory when utilizing the FRDM mass model and \cite{panov2010} fission rates. Corresponding fission fragment production rates are shown in the bottom panel. The distribution model here is ABLA07.}
\label{fission_region}
\end{figure*}

For the fissioning nuclei produced in the present r-process simulations, our fission models mainly result in fission fragments in the mass range $100 \leq A \leq 160$. Therefore, it is no surprise that the largest differences 
between the models are found around the second peak. The results obtained with the two Panov models (Fig.~\ref{ffdm13}) show a drastic underproduction of the mass region beyond the second peak ($A\approx 140-170$) by a 
factor of 10 and more, due to the dominance of the symmetric fission channel and a large number of released neutrons. 
The Kodama \& Takahashi model, in contrast, shows an overproduction of these nuclei and fails to produce a distinct second peak.
The ABLA07 model (dashed line in Fig.~\ref{ffdm24}) shows the best overall agreement with the solar r-process abundance pattern (for the chosen mass model FRDM), leading only to an underproduction of $A=140-170$ nuclei by a factor of about 3. 
\begin{figure*}[b!]
  \includegraphics[width=0.5\textwidth]{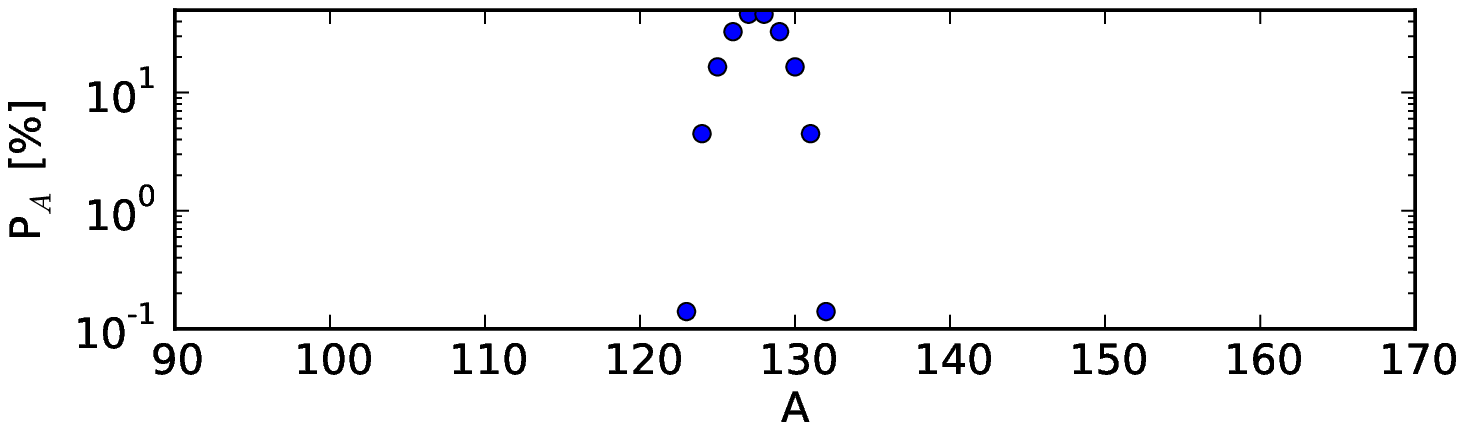}
  \label{ffd_fiss3}
  \includegraphics[width=0.5\textwidth]{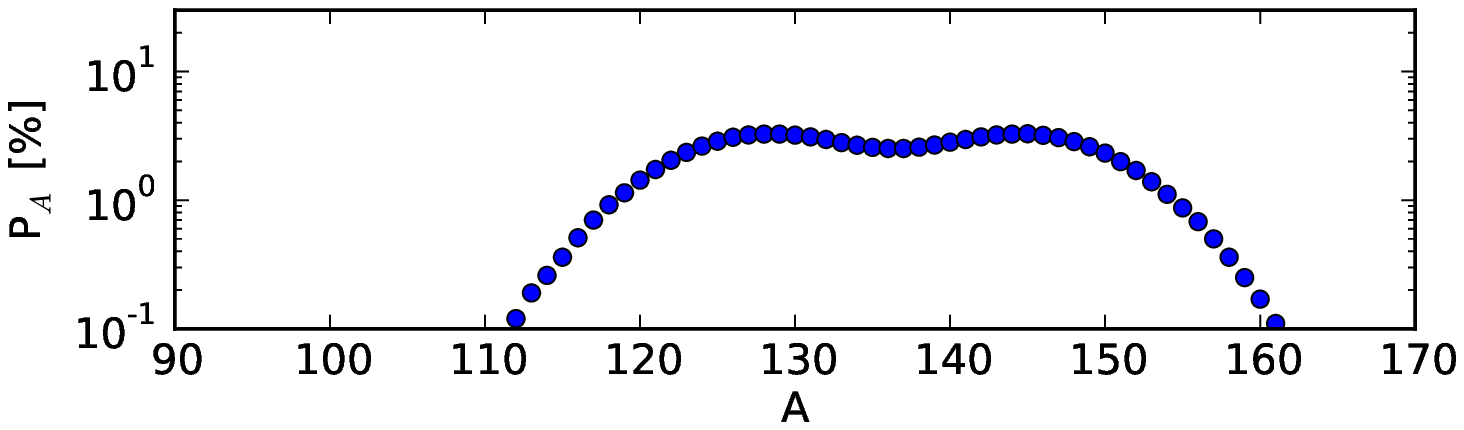}\\
  \label{ffd_fiss2}
  \includegraphics[width=0.5\textwidth]{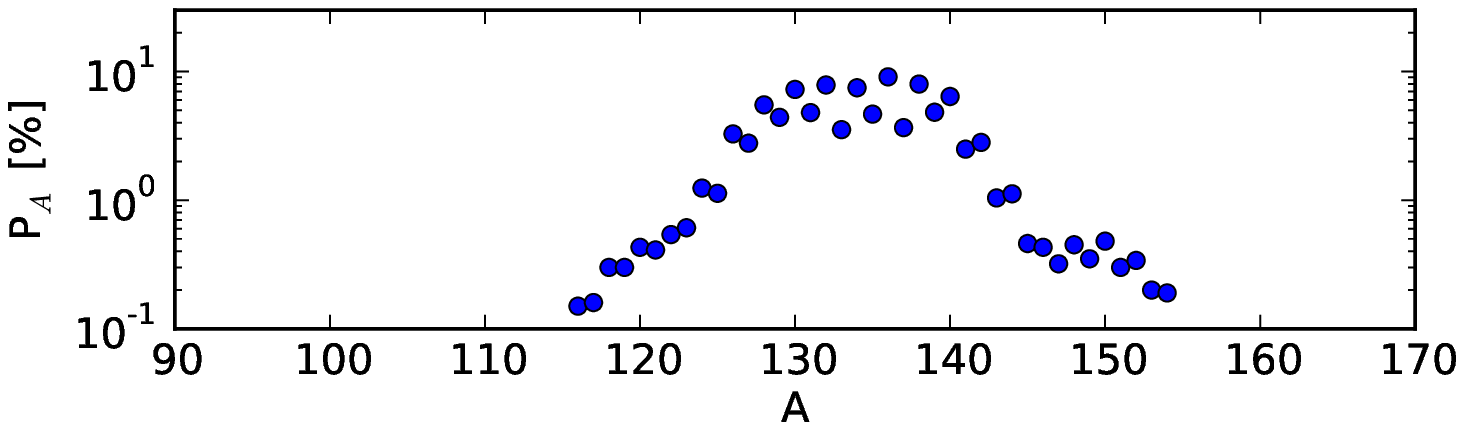}
  \label{ffd_fiss4}
  \caption{Fission fragment distributions for the models considered in our calculations, here for the case of neutron-induced fission of $^{274}$Pu (top left: \citealt{panov2008}, top right: \citealt{kodama1975}, bottom left: \citealt{kelic2008}). For this reaction \cite{panov2008} predicts 19, ABLA07 7 released fission neutrons. \cite{kodama1975} do not predict any fission neutrons. For \cite{panov2001} neutrons can be released if the fragments would lie beyond the neutron dripline. The distribution for \cite{panov2001} consists only of two products with A$_1 = 130$ and A$_2 = 144$.}
\label{ffd_pu274}
\end{figure*}
Figure~\ref{fission_region} demonstrates the importance of fission in our calculations, indicating the fission rates from two fission modes (neutron-induced and $\beta$-delayed fission) at $t = 1$s. It is obvious that the mass region with $Z=93-95$ and $N=180-186$ dominates. In Fig.~\ref{fragments_abla} we show the corresponding (combined) fragment production rates for ABLA07 in the nuclear chart. In Fig.~\ref{ffd_pu274} (and the related caption) we also provide the fission fragment distributions as a function of A as well as the number of released neutrons, for $^{274}$Pu ($Z=94$). Note that the model by \cite{kodama1975} does not lead to any neutron release and \cite{panov2008} predicts the largest number of released neutrons. It can also be seen that the predicted fragments in the Panov et al. distributions do not extend beyond $A=140$ and thus lead to the strongest underproduction in the mass range $A=140-170$. We will come back to the importance of this topic later, when we address the question whether the release of neutrons from fission dominates over $\beta$-delayed neutron emission.
\begin{figure*}[b!]
  \includegraphics[width=0.5\textwidth]{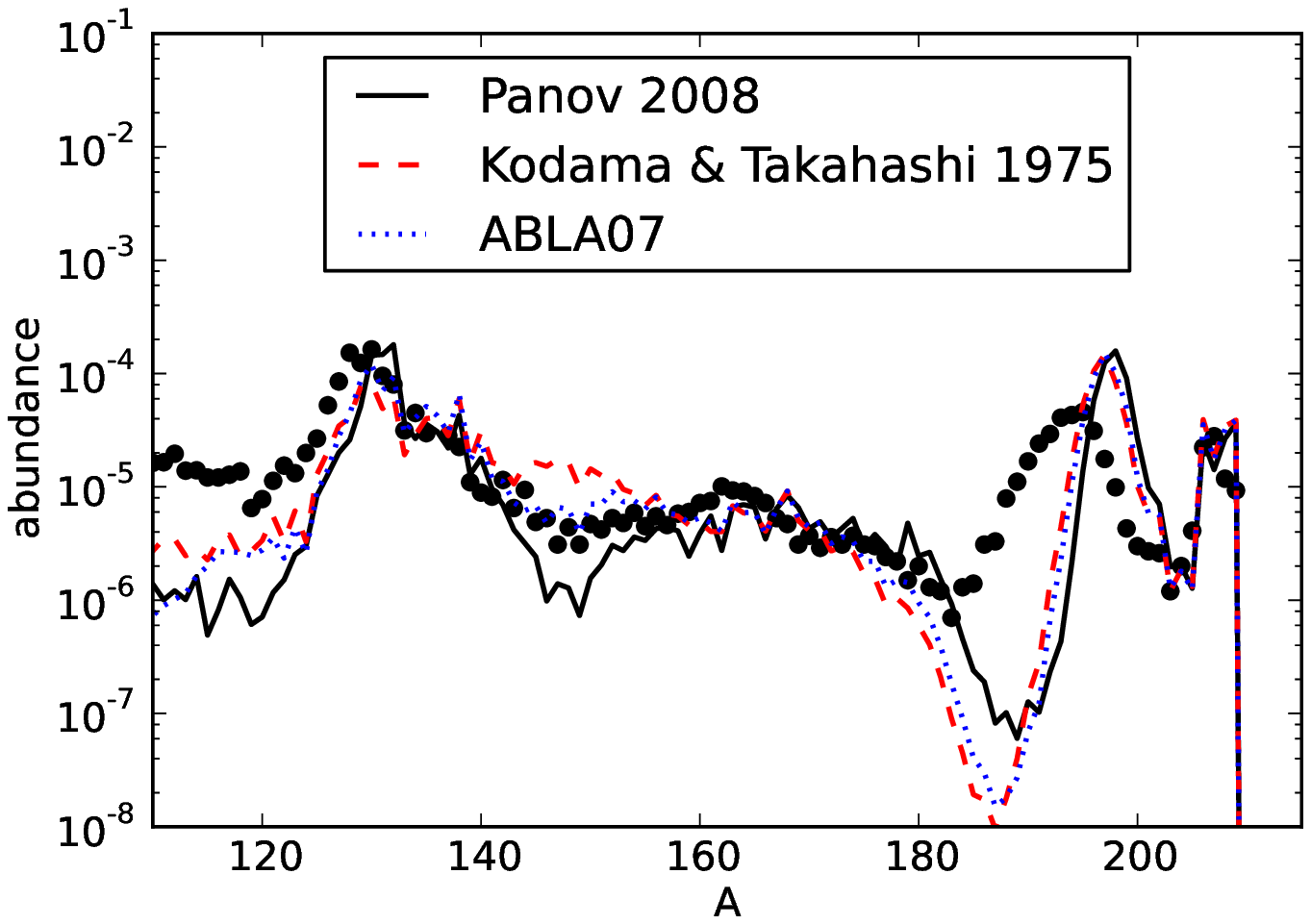}
  \label{models1a}
  \includegraphics[width=0.5\textwidth]{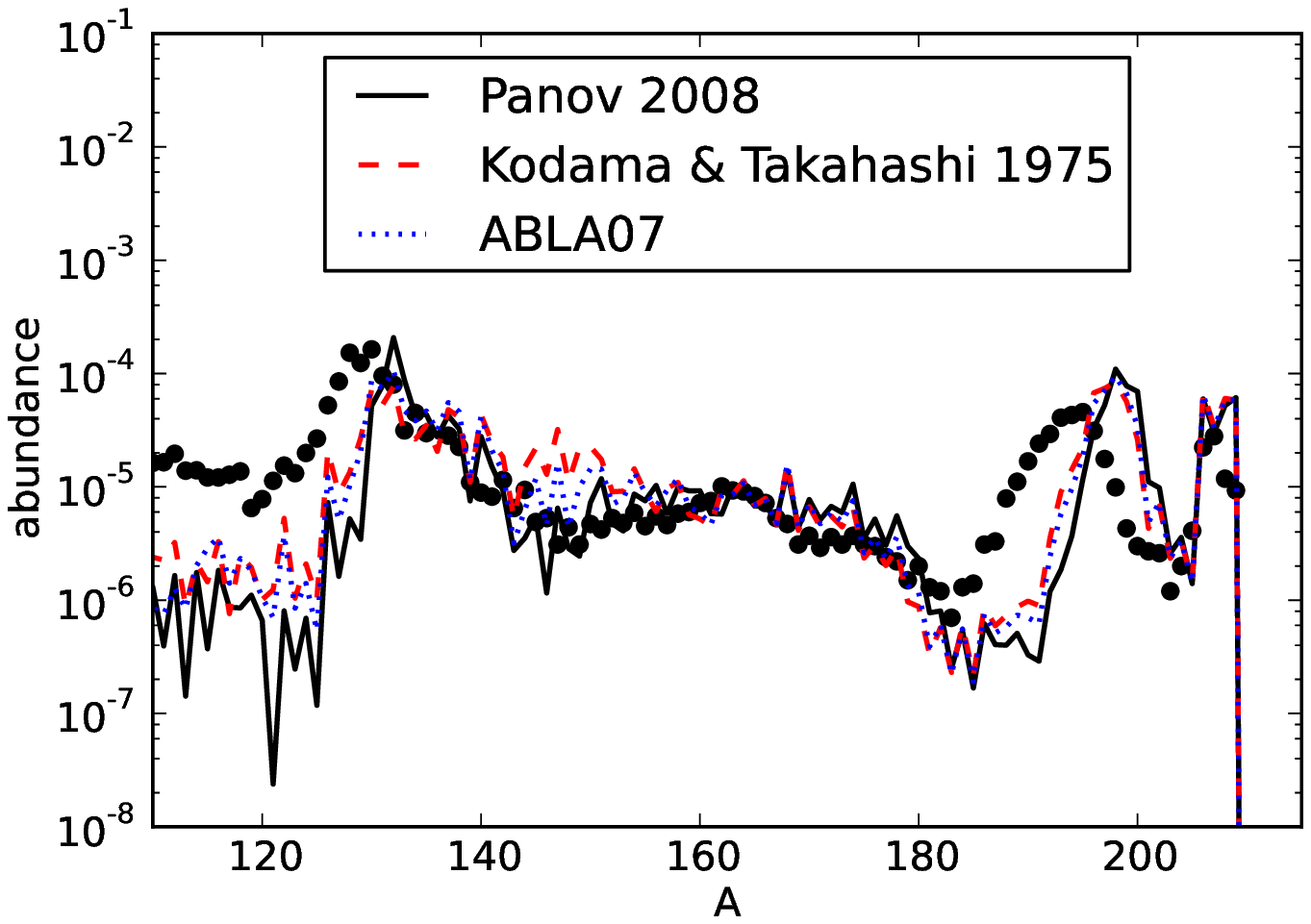}
  \label{models1b}
  \caption{Similar to Fig.~\ref{ffdm}, several fission fragment distributions are tested for the mass models ETFSI-Q (left) and 
HFB-14 (right). It can be realized that in both cases the ABLA07 fragment distribution leads to a good fit to solar r-abundances
in the mass region $A=140-170$. In addition, these mass models also avoid the still (to some extent) existing underproduction due to FRDM, apparent 
in Fig.~\ref{ffdm} also for the ABLA07 fragment distribution. The second peak in HFB-14 is slightly shifted to higher masses, a feature also seen in \cite{bauswein2013}. Whether this is due to different fission fragments or late neutron captures after fission will be discussed later.}
\label{massmodels1}
\end{figure*}
Some of the deficiencies beyond the second peak can also be attributed to the FRDM mass model, which is known to predict rather low or even negative neutron separation energies for nuclei beyond the $N=82$ shell closure around $N=90$ 
($A\sim 138$) (e.g., \citealt{meyer1992,chen1995,arconesGMP2011}). As a consequence, material is piled up in and slightly above the 
second peak, while the mass region beyond $A=140$ is underproduced. This effect might be reduced when applying the new FRDM version which is not publicly available yet \citep{moeller2012}, see e.g., \cite{kratz2014}.
Thus, in order to explore the full dependence on uncertainties due to the combination of mass models and fission fragment 
distributions, we also performed reference calculations, employing the ETFSI-Q mass model \citep{pearson1996} and the HFB-14 model
\citep{goriely2008,goriely2009} for the set of fragment distributions \cite{kodama1975}, \cite{panov2008}, and ABLA07 \citep{kelic2008}. They show less or no underproduction for $A>140$, even for the \cite{panov2008} fragment distribution.
The results are displayed in Fig.~\ref{massmodels1}.
\begin{figure*}[t!]
  \centering
  \includegraphics[width=0.6\textwidth]{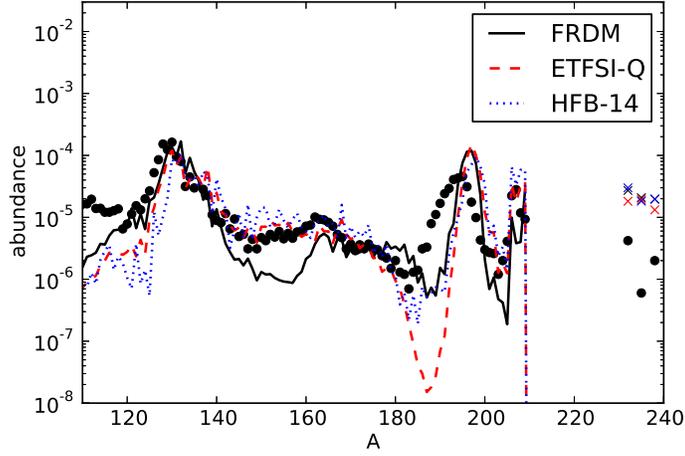}
  \caption{Comparison of nuclear mass models FRDM, ETFSI-Q, and HFB-14. The underproduction of $140 < A < 160$ nuclei apparent in the FRDM model does not occur in the ETFSI-Q or HFB-14 model cases. The fission fragment distribution model used here is ABLA07.}
\label{massmodels}
\end{figure*}
A comparison of all different mass models with the fragment distribution ABLA07 is shown in  Fig.~\ref{massmodels}. ETFSI-Q suffers from a sudden drop of the neutron separation energy for $A \approx 140$, causing the formation of a small peak 
around this mass number. The distinctive trough in the ETFSI-Q abundance distribution before the third peak was
subject of a detailed discussion in \citet{arconesGMP2011}. While the extent of the underproduction in the mass range $140-160$
is due to a combination of the fission fragment distribution and the mass model used (see also Fig.~\ref{massmodels1}), the results for all mass models utilized here show a shift of the third peak to higher mass numbers by up to 5 units, which will be a topic of the following subsections.
\begin{figure*}[p!]
%  \centering
  \includegraphics[width=0.55\textwidth]{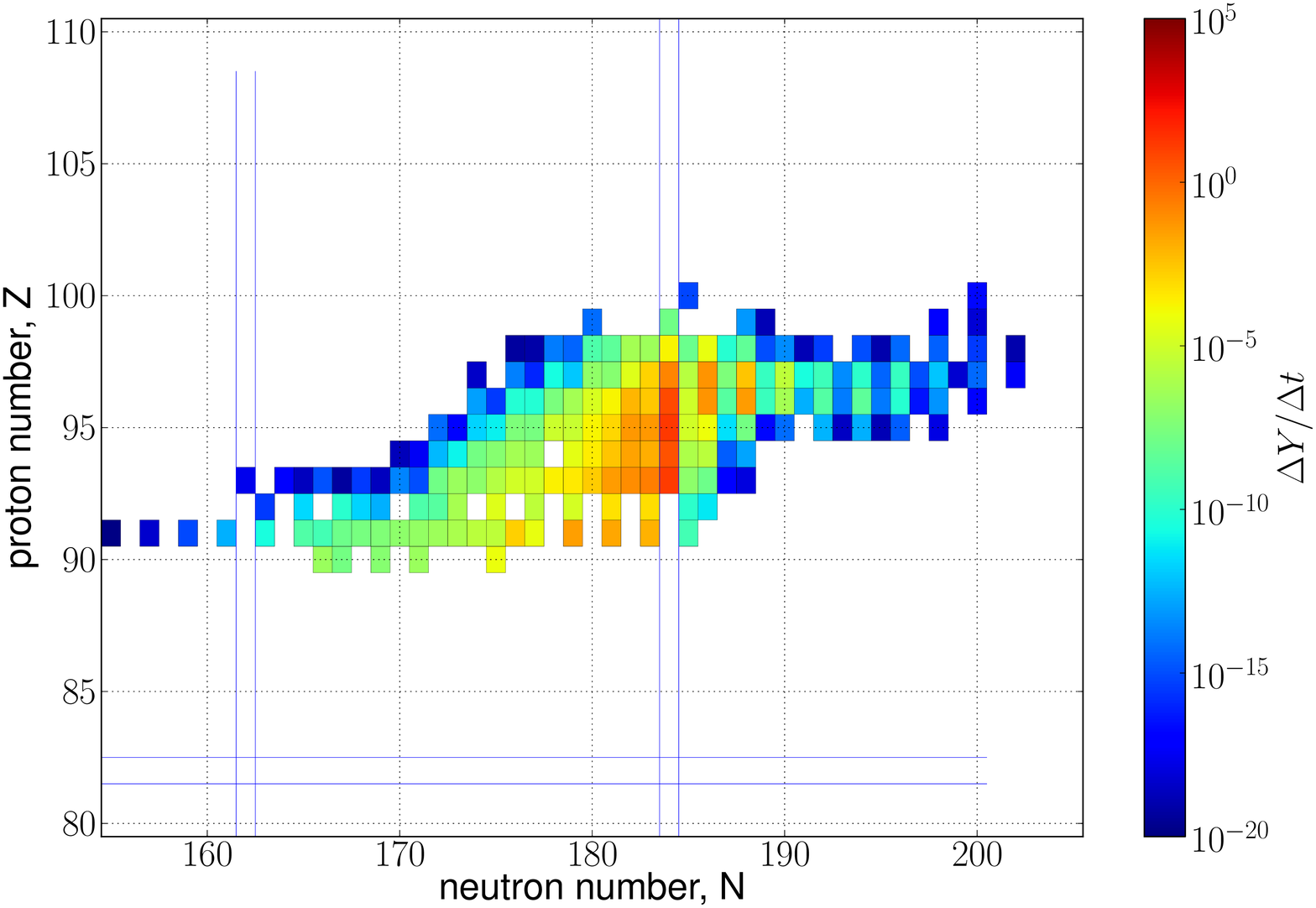}
  \label{hfb_parents_beta}
  \includegraphics[width=0.55\textwidth]{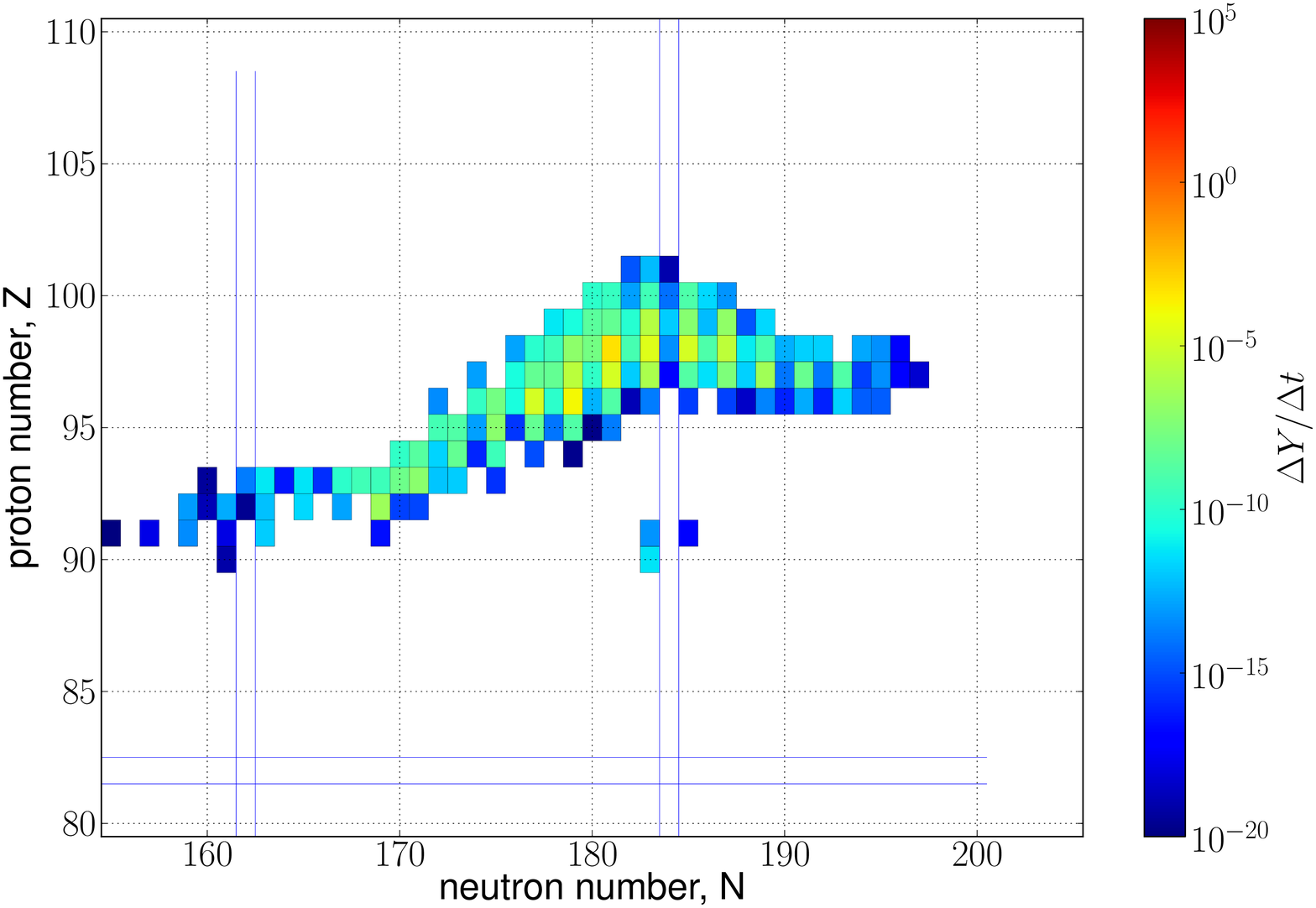}\\
  \label{hfb_parents_n}
  \hspace{1.5cm}
  \includegraphics[width=0.8\textwidth]{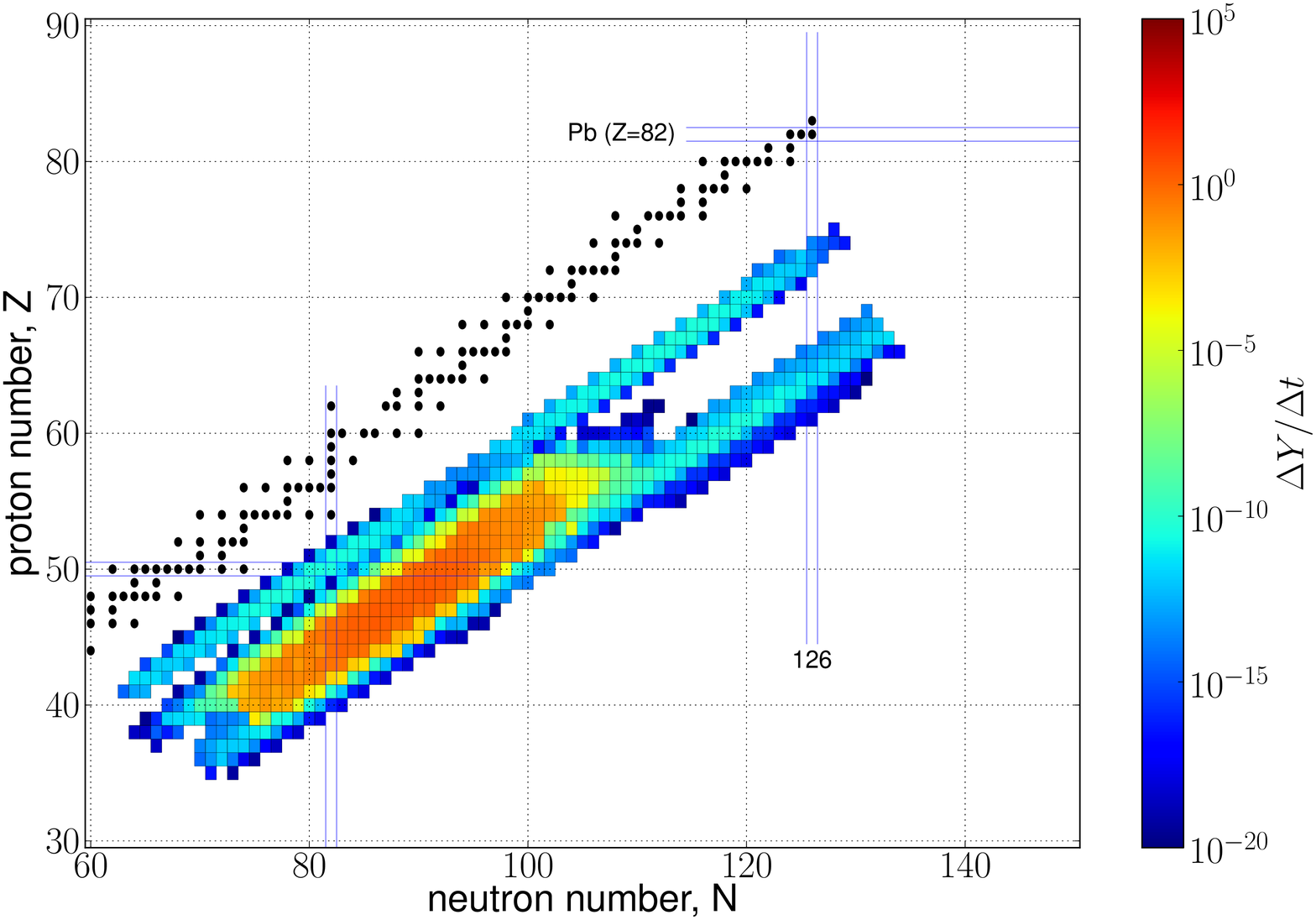}
  \label{hfb_fragments}
  \caption{Same as Fig.~\ref{fission_region} (top left: $\beta$-delayed fission; top right: neutron-induced fission; bottom: fragment production rates), but for the HFB-14 mass model.}
\label{hfb_parents}
\end{figure*}
While the position of the third peak is similar for all the mass models considered here, the abundance patterns around the second peak and the rare-earth peak show some diversity. For these mass regions, the final abundances are strongly influenced by fission close to the freeze-out and also possible final neutron captures thereafter. Therefore, different final abundance patterns can be an indicator of different fission progenitors. Figure~\ref{hfb_parents} shows the predominant fission reactions at the time of freeze-out for the HFB-14 model. A comparison with Figure~\ref{fission_region} reveals that for the HFB-14 model the fission close to freeze-out tends to happen at higher mass numbers (up to $A = 300$), while for the FRDM model the fission parent with the highest mass is found at $A=287$. As a consequence, fragments with higher mass can be produced (Fig.~\ref{hfb_fragments}). However, the bulk of fragments lies between $A = 125$ and $A = 155$, very similar to the FRDM case. Therefore, the aforementioned shift of the second peak in the HFB model calculations cannot be due to the fission fragment distribution lacking fragments with mass numbers at the lower flank of the second peak. The main cause must be reactions occurring after fission, which will be discussed in the following section.
% and ultimately in the final abundance distribution around and above the second peak, with HFB-14 failing to reproduce the low-mass flank of the second peak (see Fig.~\ref{models1b}).}
%\subsection{The Position of the Third Peak}
\subsection{The Impact of Late Neutron Captures}
\label{thirdpeak}
\noindent In our NSM calculations, the third peak is shifted towards higher mass numbers compared to the solar values (Figures~\ref{ffdm}~,~\ref{massmodels1},~\&~\ref{massmodels}), regardless of the nuclear mass model utilized in the present investigation. This phenomenon has appeared in various calculations of NSMs 
before \citep{freiburghaus1999,metzger2010b,roberts2011,korobkin2012,goriely2013}. We find that the position of the 
third peak in the final abundances is strongly dependent on the characteristics of the conditions encountered during/after the r-process freeze-out that are characterized by a steep decline in neutron density and a fast increase in the timescales for neutron captures and photodissociations, leading to different stages (timescales): (1) freeze-out from an (n,$\gamma$)-($\gamma$,n) equilibrium, (2) almost complete depletion of free neutrons ($Y_n / Y_{seed} \leq 1$), and (3) the final abundance distribution. In the following, we use the term \textit{freeze-out} in the context of definition (1).
\begin{figure*}[t!]
  \centering
  \includegraphics[width=\textwidth]{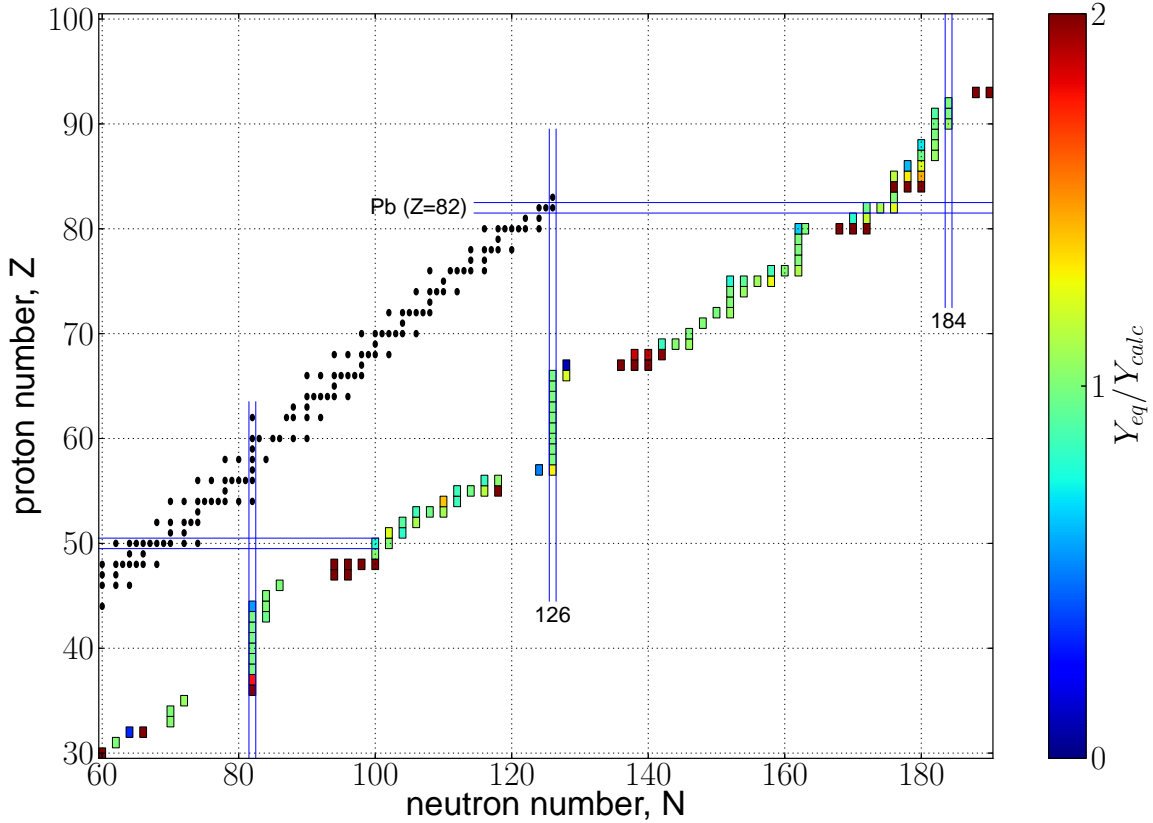}
  \caption{Comparison of abundances from our calculations with (n,$\gamma$)-($\gamma$,n) equilibrium abundances on the r-process path for the FRDM mass model. The colours show the factor $Y_{eq}/Y_{calc}$. Only the most abundant nuclei are shown for each isotopic chain. See text for details.}
\label{ab_factor}
\end{figure*}
Figure~\ref{ab_factor} shows a comparison of our abundances on the r-process path resulting from detailed nucleosynthesis calculations at $t=1$~s for the FRDM mass model with those which would result from an (n,$\gamma$)-($\gamma$,n) equilibrium in each isotopic chain (as first discussed by \citealt{seeger1965}) for the temperature and neutron density at that time ($T=9.5 \times 10^8$~K, $n_n = 7.44 \times 10^{26}~cm^{-3}$). The plot displays the most abundant nuclei in each isotopic chain, i.e., those on the r-process path. The colours indicate the factor between the equilibrium abundances and the abundances in our calculation. The highest discrepancies can be observed around $N=100$ and $N=140$, but only few nuclei show a factor larger than 2. This leads to the conclusion that at this time the r-process still proceeds in (n,$\gamma$)-($\gamma$,n) equilibrium with (n,$\gamma$) and ($\gamma$,n) timescales much shorter than $\beta$-decays, characteristic of a hot r-process.
\begin{figure*}[t!]
  \centering
  \includegraphics[width=0.6\textwidth]{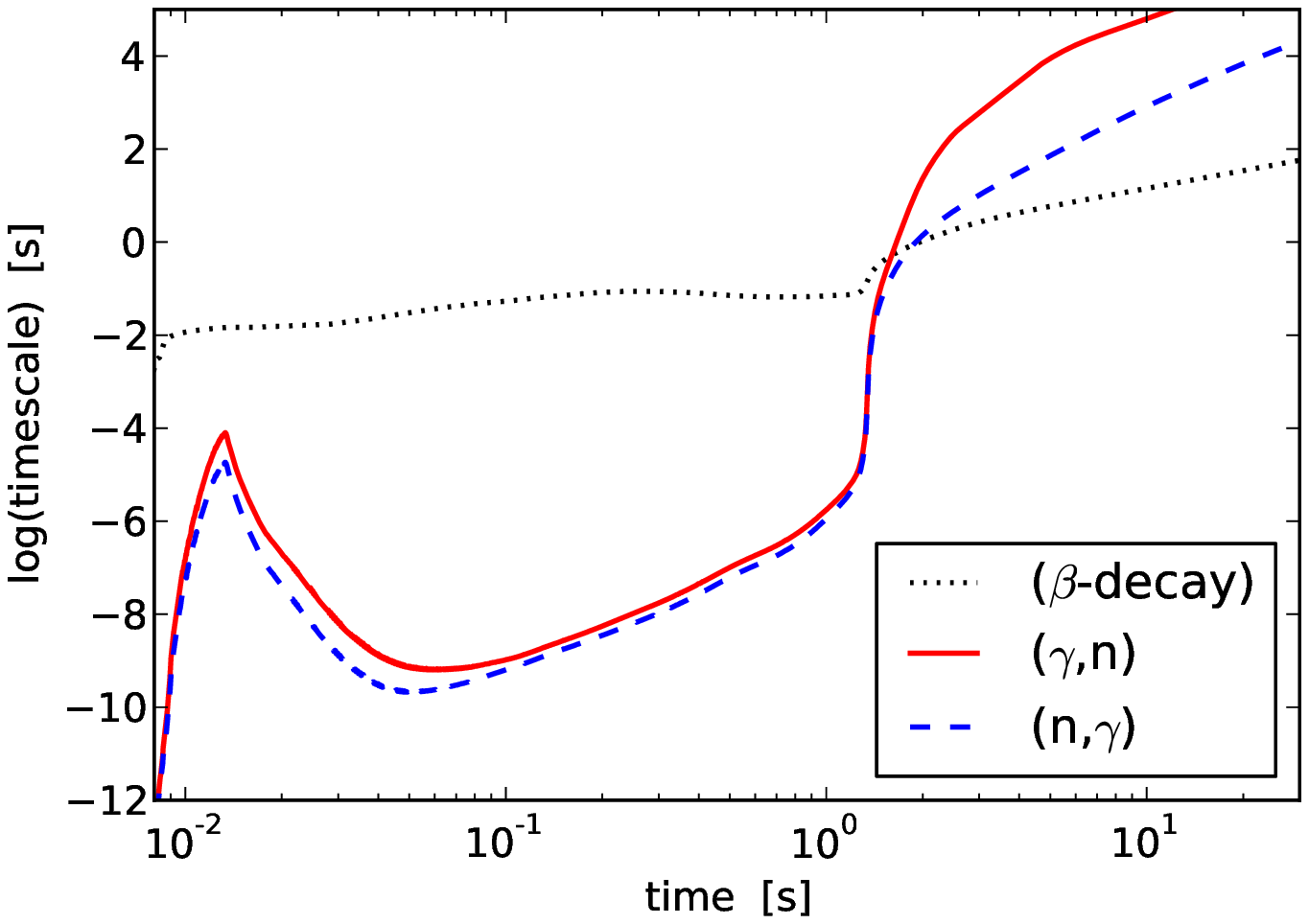}
  \caption{Averaged reaction timescales vs. time for $\beta$-decays, ($\gamma$,n) reactions, and (n,$\gamma$) reactions for one trajectory with the FRDM mass model.}
\label{timescales}
\end{figure*}
\begin{figure*}[p!]
  \centering
  \includegraphics[width=0.8\textwidth]{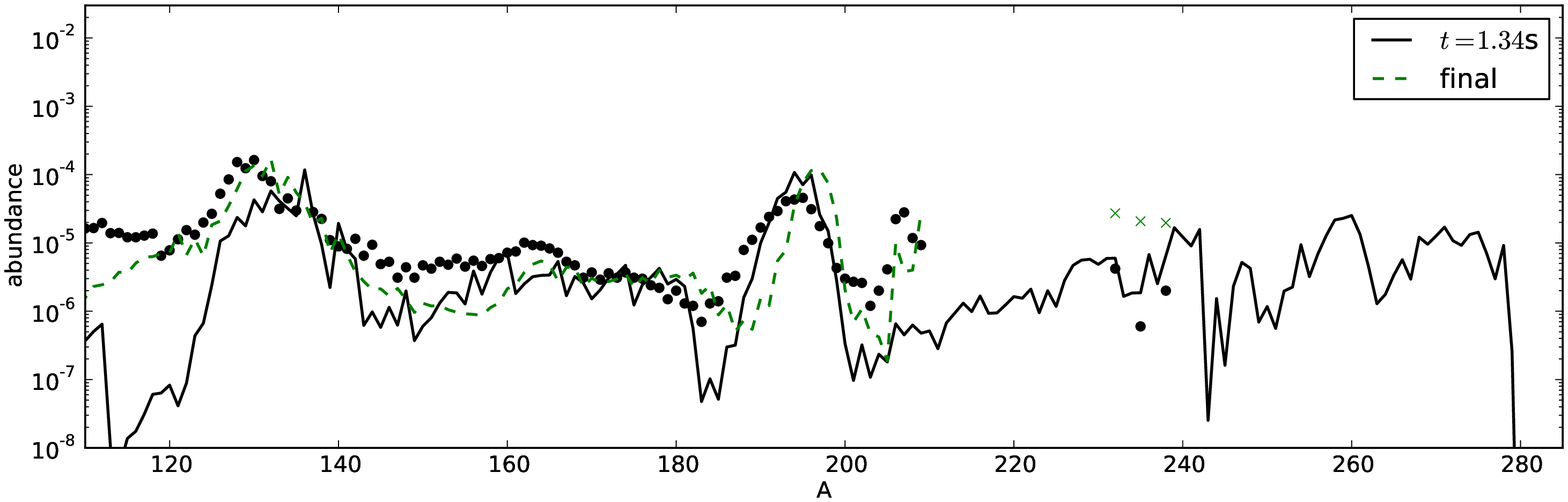}\\
  \label{freezeandfinal_frdm}
  \includegraphics[width=0.8\textwidth]{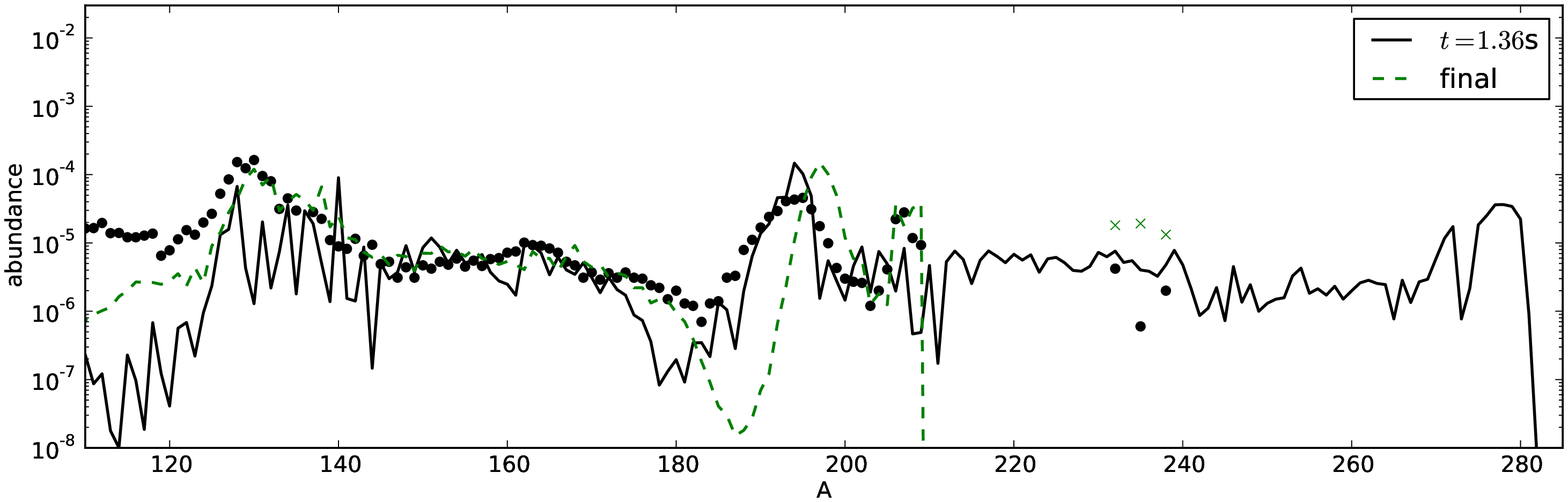}\\
  \label{freezeandfinal_etfsiq}
  \includegraphics[width=0.8\textwidth]{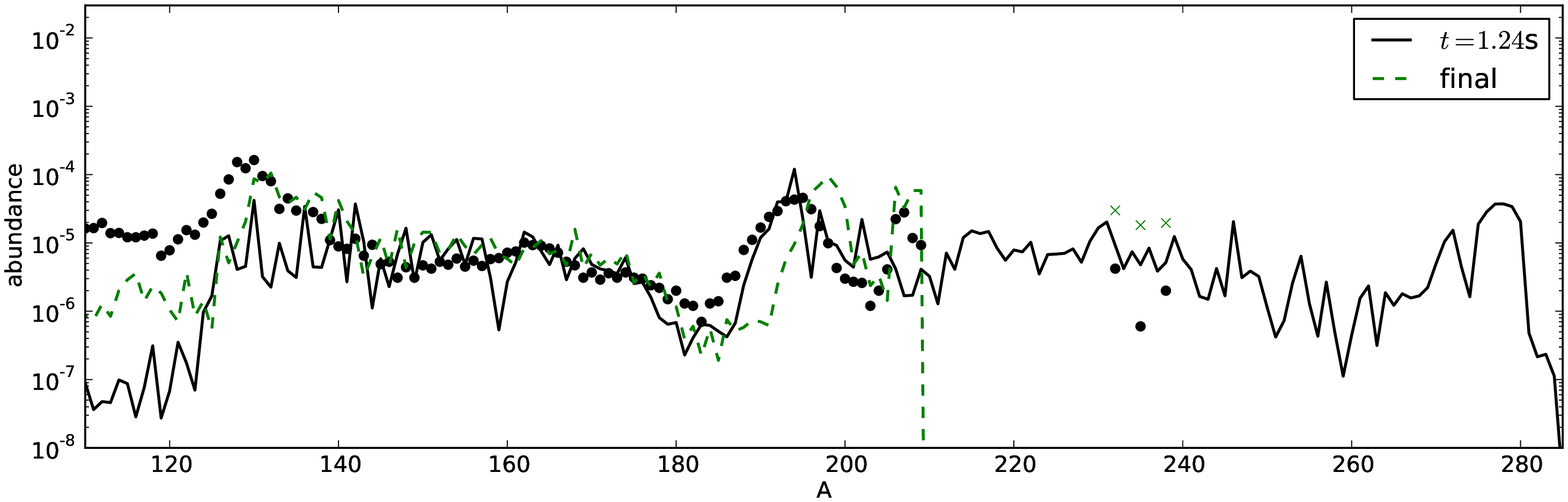}
  \label{freezeandfinal_etfsi}
  \caption{Top panel: Second and third peak abundances at the time of r-process freeze-out (where $\tau_{(\gamma,n)} > \tau_{(n,\gamma)}$ or $ t = 1.34$~s; solid line) compared to the final abundances (dashed line) for one trajectory employing the FRDM mass model. Middle panel: Same for the ETFSI-Q mass model, where the freeze-out occurs at $t=1.36$~s. Bottom panel: Same for the HFB-14 mass model. Here the freeze-out occurs at $t = 1.24$~s. Notice that the third peak position is still consistent with the solar r-abundances at freeze-out, but that for all mass models a shift takes place afterwards. For none of the mass models the features of the second peak at freeze-out are perfect. However, the FRDM and ETFSI-Q models show a decent agreement for the final abundances, while for the HFB-14 mass model the second peak is shifted for the time at freeze-out as well as for the final abundances.}
\label{freezeandfinal}
\end{figure*}

\begin{figure*}[h!]
  \centering
  \includegraphics[width=0.6\textwidth]{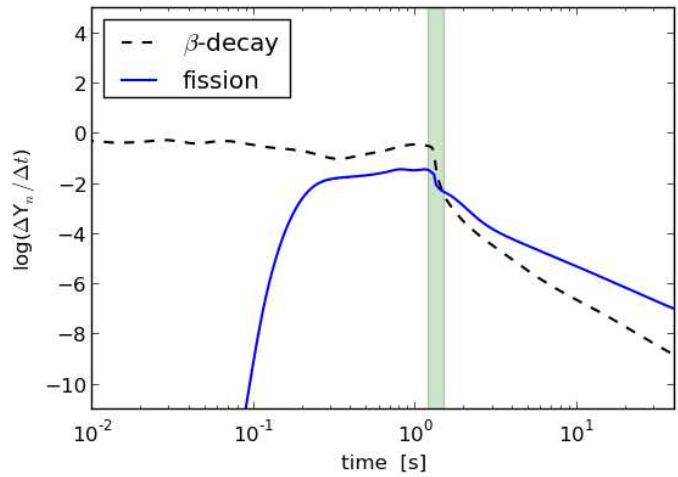}
  \caption{Released neutrons due to fission and $\beta$-delayed neutrons vs. time. The plotted quantity is the neutron production rate (per second). After the time of freeze-out (shaded area) fission neutrons dominate over $\beta$-delayed neutrons.}
\label{released_neutrons}
\end{figure*}

\begin{figure*}[h!]
%  \centering
  \includegraphics[width=0.5\textwidth]{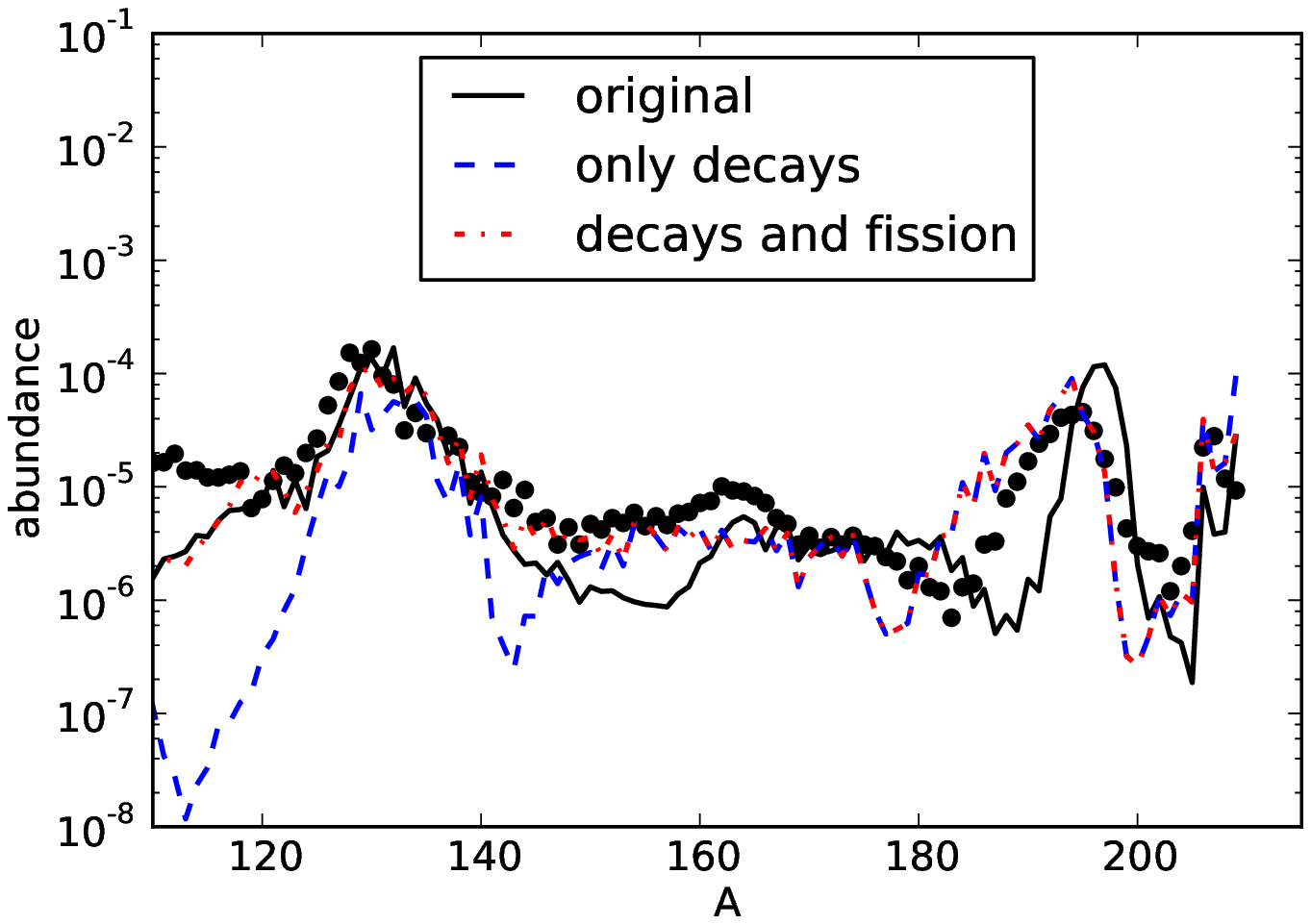}
  \label{onlydecays_frdm}
  \includegraphics[width=0.5\textwidth]{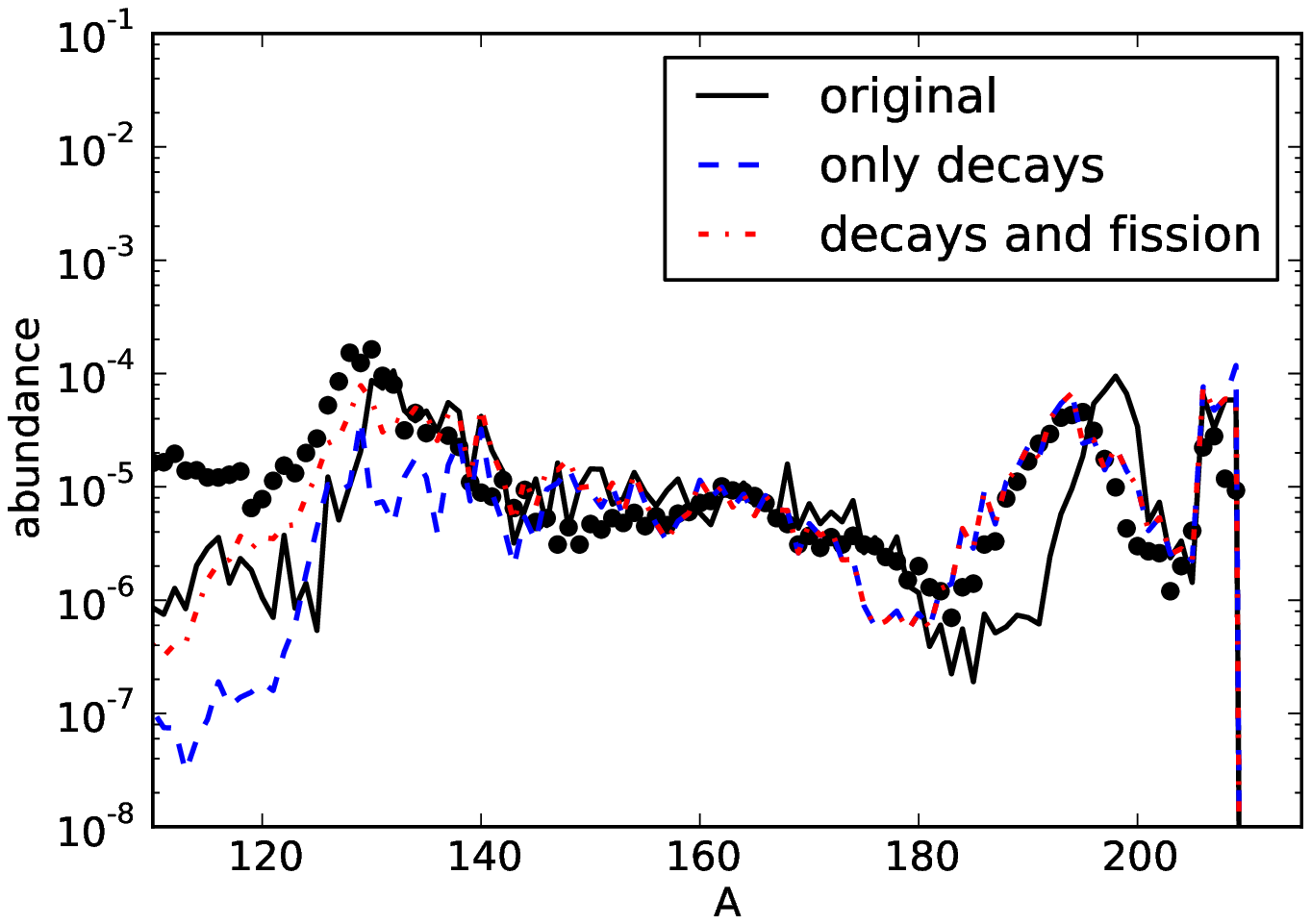}
  \label{onlydecays_hfb}
  \caption{Final abundance distribution for cases where only certain types of reactions are allowed to proceed after freeze-out (dashed line: only decays except for fission; dot-and-dashed line: decays including fission) for FRDM (left) and HFB-14 (right). The solid line represents the original calculation where neutron captures are also allowed after freeze-out. All cases use ABLA07.}
\label{onlydecays}
\end{figure*}

This changes at $t=1.34$~s (see Fig.~\ref{timescales}), when the timescales for neutron capture and photodissociation become larger than the $\beta$-decay timescale.
Here both reaction timescales become longer than $\beta$-decays, and also neutron capture wins against photodissociations. Note that the timescales of $\beta$-decay also become larger as the material moves closer to stability. As can be seen in Fig.~\ref{timescales}, there is a short period after the freeze-out where (n,$\gamma$) dominates over both ($\gamma$,n) and $\beta$-decay.
Figure~\ref{freezeandfinal} shows the second and third peak abundances at r-process freeze-out and the final abundances for a representative trajectory for the FRDM, ETFSI-Q, and HFB-14 mass models. It is evident that the position of the third peak is still in line with the solar peak at freeze-out, but is shifted thereafter for all mass models. The position of the second peak behaves differently.
For all mass models, the final abundances for $A < 120$ nuclei are higher than the abundances at freeze-out, because fission fragments with these mass numbers are still produced after freeze-out. Nevertheless for the HFB-14 model the (final) second peak seems shifted to higher mass numbers, similar to its position at freeze-out. This might indicate that, for the astrophysical conditions encountered here, this mass model leads to a path running too close to stability.

The shift in the third peak as described above is a generic feature in our NSM calculations. It is caused by the continuous supply of neutrons from the fissioning of material above $A \simeq 240$. Fig.~\ref{released_neutrons} shows that after the freeze-out the release of neutrons from fission dominates over $\beta$-delayed neutrons. 

To further illustrate the importance of fission neutrons after the freeze-out, we have run several calculations with both FRDM and HFB-14 where we have switched off certain types of reactions after the freeze-out. (1) The dashed lines in Figure~\ref{onlydecays} (labelled ``only decays'') represent the cases where only decay reactions are allowed after the (n,$\gamma$)-($\gamma$,n) freeze-out (without fission). In this artificially created scenario the only possibility for nuclei after the freeze-out is to decay to stability, without the option to fission or capture neutrons. In fact, a small shift of the third peak to lower mass numbers can be observed during this phase (compare Fig.~\ref{onlydecays} \& \ref{freezeandfinal}), as $\beta$-delayed neutrons cause the average mass number to decrease. In addition, since fission is not allowed either, the second peak consists of just the material that was present there at freeze-out, but the composition is (slightly) modified due to the combined effects of $\beta$-decays and $\beta$-delayed neutrons. (2) If we also allow for fission in addition to the decay reactions (dot-and-dashed lines in Fig.~\ref{onlydecays}), the second peak is nicely reproduced by fission fragments for both mass models and the third peak is still not affected. (3) However, a notable difference between the two mass models can be seen for the final abundance distribution including also final neutron captures (denoted as ``original'' in Fig.~\ref{onlydecays}), indicating that for HFB not only the position of the third peak is influenced by late neutron captures, but also the position of the second peak. On the other hand, the behaviour is reversed for the mass region $140 < A < 160$, where large deviations can be observed compared to the original calculation for FRDM, since in the original case neutron captures move material up to higher masses, creating the underproduction we have discussed in chapter~\ref{ffd}. This indicates that when also neutron captures and all other reactions are permitted after freeze-out (i.e., the original calculation), major changes in the abundance pattern can still occur. The third peak moves to higher masses for all mass models discussed here. In the HFB case, the second peak moves to the position it had at freeze-out, resembling abundance features resulting from an r-process path too close to stability for the astrophysical conditions encountered here.

It can be seen in Figures~\ref{ffdm}~,~\ref{massmodels1}~\&~\ref{freezeandfinal} that the shift of the third peak is indeed related to the amount of released neutrons. The shift is smallest for the \citet{kodama1975} model, which does not assume any neutron emission during fission, and largest for the \citet{panov2008} model, which assumes the largest amount of neutrons produced.
It should be noted, however, that the differences are smaller than expected from the numbers of released neutrons as in addition to the neutrons directly released during fission also those released from the neutron-rich fission fragments can be important as they move back to the r-process path. These in turn depend on the fission yields used.

Since the position of the third peak does not coincide with the third peak of the solar abundance pattern, we explore in the following under which conditions such a shift to larger mass numbers can be avoided. In a first test we artificially increase the temperature throughout the expansion by setting the heating efficiency parameter $\epsilon_{th} = 0.9$ instead of our default value of $0.5$ (see section~\ref{method}). This change does not affect the final abundance distribution significantly, in particular the position of the third peak, because more vigorous heating simply prolongs the (n,$\gamma$)-($\gamma$,n) equilibrium until the temperature drops to a similar value due to expansion. Therefore, the r-process freeze-out happens later, at a temperature that is comparable to the reference case ($\epsilon_{th} = 0.5$). Our finding that the exact value of the heating efficiency parameter $\epsilon_{th}$ does not greatly affect the final abundances, provided that it is above some threshold value, is in agreement with \citet{korobkin2012}.

We have further explored the effect of modified neutron capture rates. Slower rates could arise as the statistical model might not be applicable for small neutron separation energies S$_n$ and not sufficiently high level densities in the compound nucleus. Faster rates could be attributed to the rising importance of direct capture contribution far from stability \citep{mathews1983,rauscher2011}.
We realized that artificially varying the neutron capture rates across the nuclear chart does not have an effect on the position of the third peak. However, some minor local effects on the final abundance distribution can be observed. Reduced rates slow down the reaction flux and, as a consequence, lead to a reduced underproduction of $140< A <165$ and a slight overproduction of $180< A <190$ nuclei, the former due to fission fragments, the latter caused by $S_n$ predictions of the FRDM mass model. Accelerating the rates has an opposite, but still minor effect.

\begin{figure*}[p!]
%  \centering
  \includegraphics[width=0.5\textwidth]{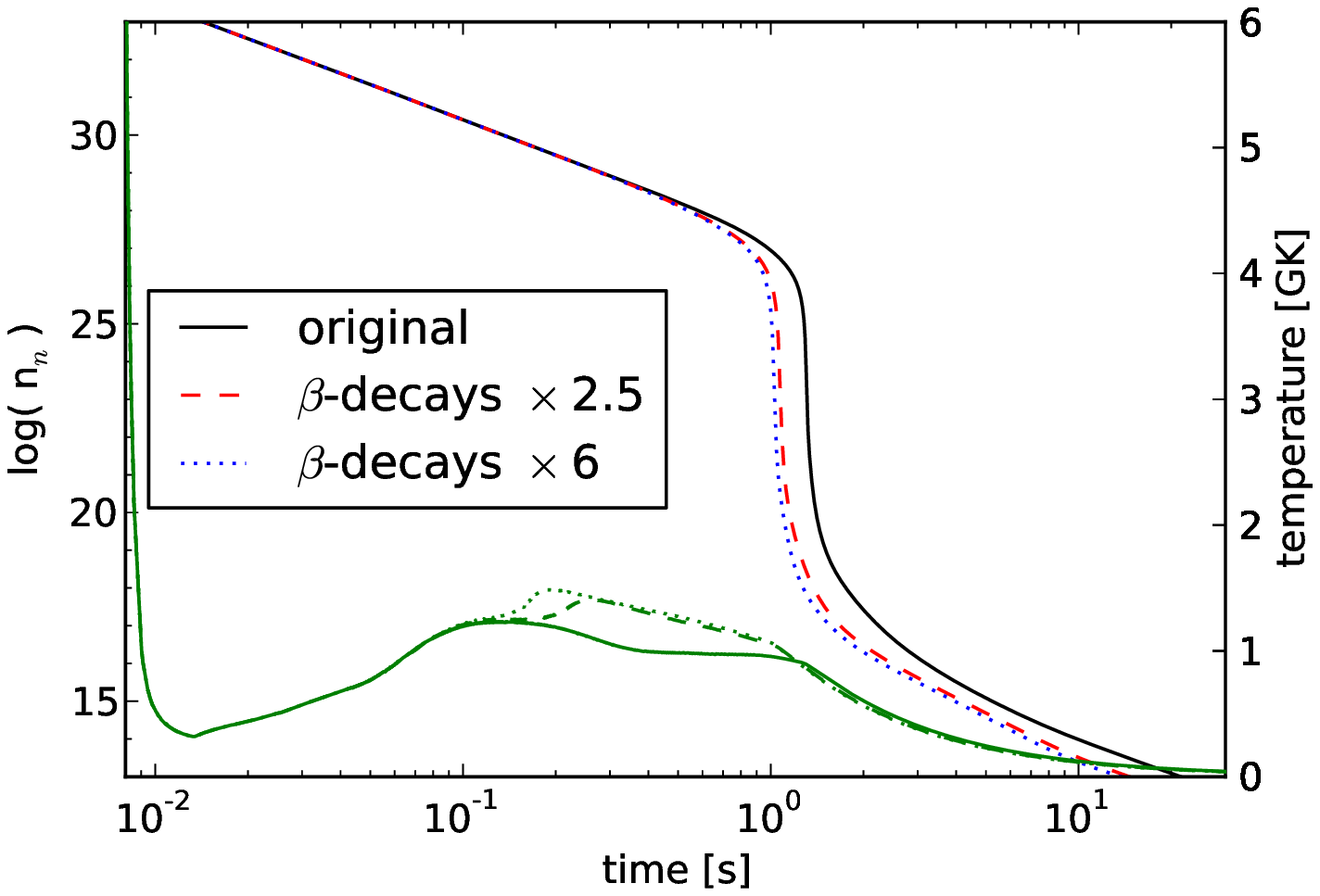}
  \label{frdm_nn}
  \includegraphics[width=0.5\textwidth]{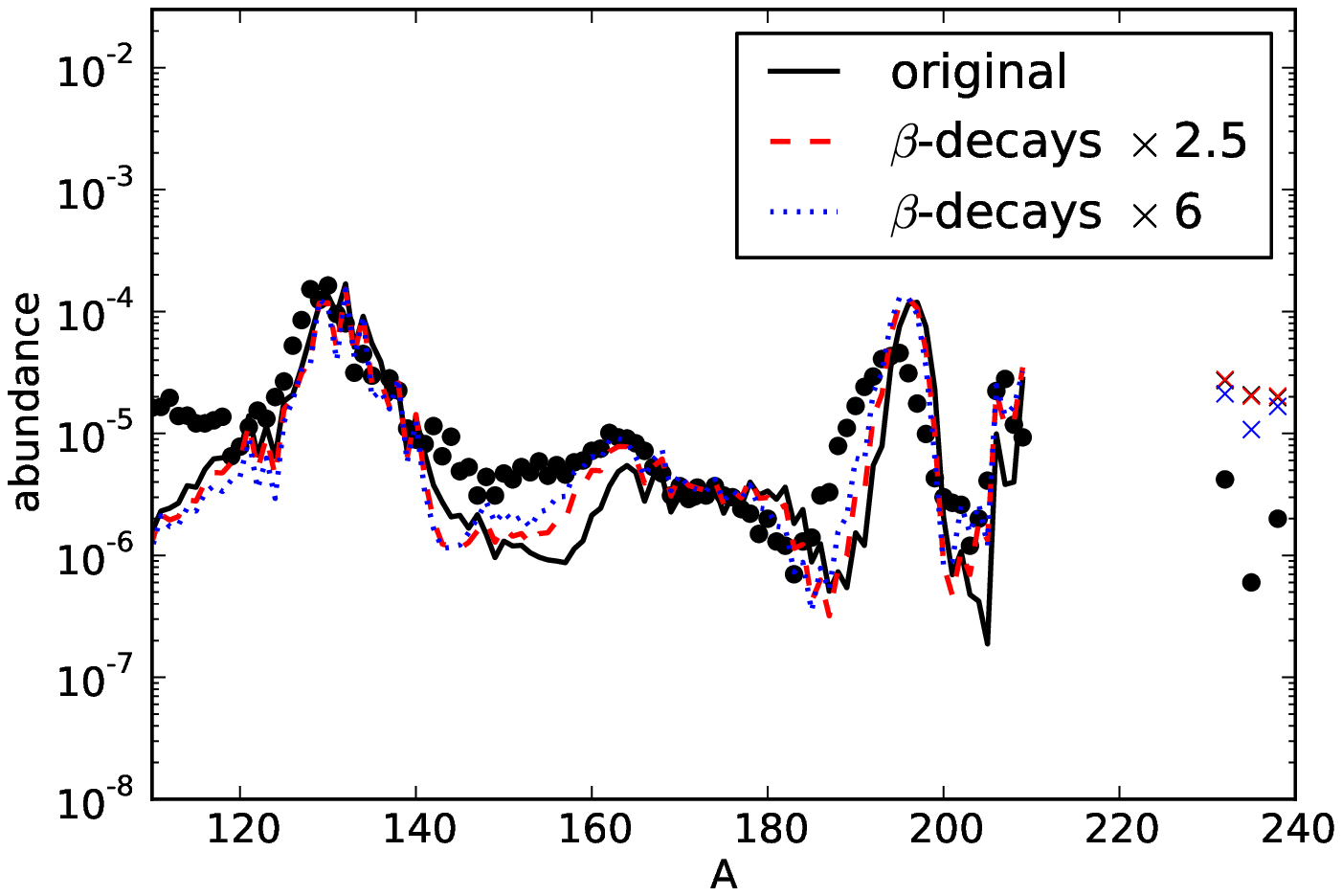}\\
  \label{fastbeta_abun}
  \includegraphics[width=0.5\textwidth]{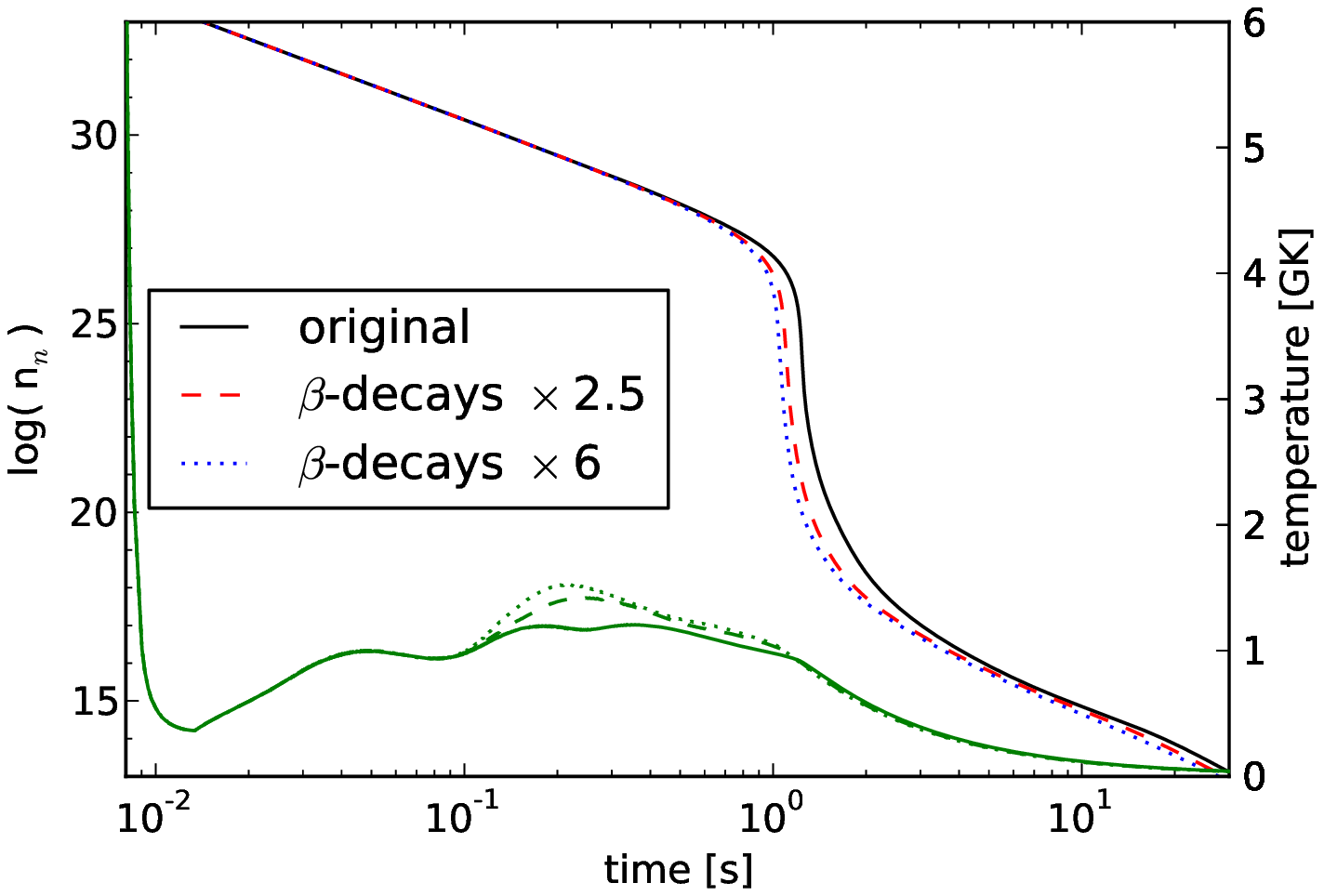}
  \label{etfsi_nn}
  \includegraphics[width=0.5\textwidth]{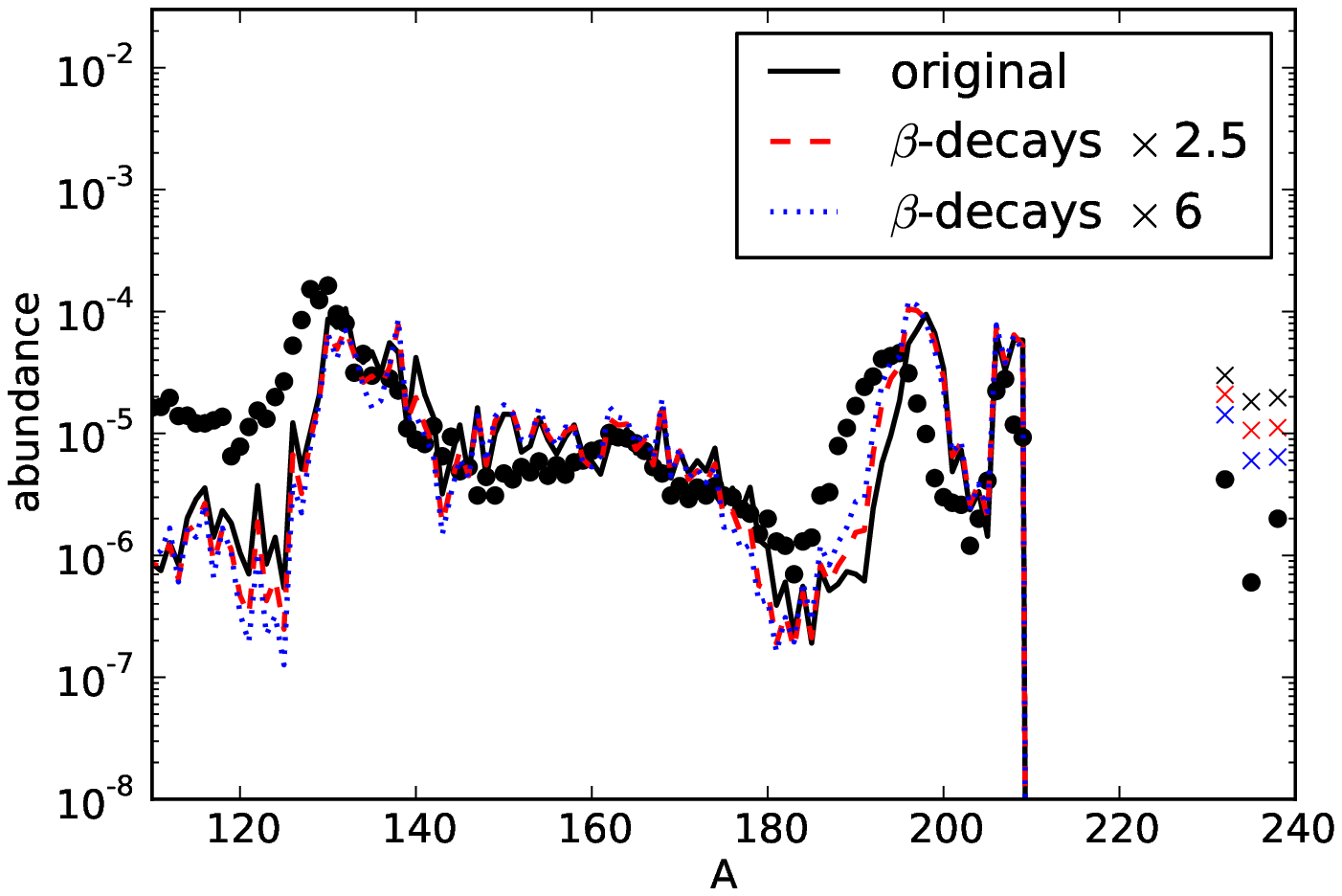}\\
  \label{fastbeta_etfsi_abun}
  \caption{Top left: neutron density ($n_n$) and temperature (green lines in the bottom part of the graphs; the linestyles correspond to the individual calculations). Top right: final abundances for a NSM calculation with artificially acccelerated $\beta$-decays (dashed and dotted line) compared to the original calculation (solid line) with the FRDM nuclear mass model. The calculations were repeated using the HFB-14 model (bottom panels). Here we use the ABLA07 fission fragment distribution model \citep{kelic2008}. See text for further explanations.}
\label{fastbeta}
\end{figure*}

As the shift is related to the continuous supply of neutrons from fission of heavy nuclei, any mechanism that affects the timescale for this supply can potentially influence the position of the third r-process peak. As an example, we have artificially increased all the $\beta$-decay rates for nuclei with $Z > 80$ (which corresponds roughly to $A>220$) by exploratory factors of 2.5 and 6, respectively. This change of rates has been motivated by recent
calculations \citep{panov2015,marketin2015}.
These latest predictions underline that especially the heavy nuclei with $Z>80$ may have shorter half-lives
by a factor of about 10. This is exactly the mass range tested in the present calculations. 
The results are shown in Figure~\ref{fastbeta} for the example of one trajectory with two nuclear mass models (FRDM and HFB). As a consequence of the increased $\beta$-decay rates, the reaction flux for the heavy nuclei is accelerated, which increases both the heating rate and the temperature at around $0.1~$s in the calculation (Figures~\ref{frdm_nn}~\&~\ref{etfsi_nn}). Additionally, the release of neutrons by fission of heavy nuclei is accelerated, providing neutrons before freeze-out (when the third r-process peak is still located close to solar values). The evolution after freeze-out proceeds faster and consequently the period of time where a combination of neutron captures and $\beta$-decays can move nuclei to higher mass numbers becomes shorter. As a consequence, the shift in the third r-process peak is reduced.

We have also tested the effect of an overall increase of (experimentally unknown) $\beta$-decay rates by constant factors across the nuclear chart. In this case the effect discussed above vanishes again, as the matter flux feeding the abundance of fissioning nuclei continues on a faster pace and thus leads to an extended release of fission neutrons.

In summary, neutron capture after (n,$\gamma$)-($\gamma$,n) freeze-out changes composition features which originate from classical r-process patterns related to an r-process path at a given neutron separation energy. This can be realized by combining the findings in Figures~\ref{freezeandfinal},~\ref{onlydecays},~\&~\ref{fastbeta}, but is complicated by a complex interaction of freeze-out, final neutron captures, plus the feeding due to fission (combined with neutron release). 
Fig.~\ref{freezeandfinal} demonstrates (for all mass models) that the third peak is shifted to higher masses during/after freeze-out, caused by the final neutron captures from neutrons which are released during fission of the heaviest nuclei in the final phases of nucleosynthesis (whereas the third peak is still located at the correct position in the last moments when (n,$\gamma$)-($\gamma$,n)-equilibrium holds). 
This feature is underlined by the results of Fig.~\ref{fastbeta}, which show the effect of accelerating the beta-decays of the heaviest nuclei ($Z>80$), i.e., accelerating the feeding of fission parents, which causes fission (and the related neutron release) to occur at different phases (before/during/after) of the freeze-out. An early neutron release (coming with the fastest $\beta$-decays of heavy nuclei) still tends to permit (n,$\gamma$)-($\gamma$,n)-equilibrium and reduces the effect of late neutron capture, although the effect is not sufficient to prevent the move of the third peak completely.
We see a similar effect in the $A=140-160$ mass region for the FRDM mass model, slowing down the movement of matter to heavier nuclei and partially avoiding the trough which appears in the final abundance pattern for the original calculation with unchanged nuclear input.
A more complex behavior causes the final abundance pattern of the second r-process peak with a complex interaction of fission feeding and final neutron processing. In Fig.~\ref{onlydecays} we see that for the FRDM as well as the HFB mass model (when utilizing ABLA07) we have an almost perfect fragment distribution in order to reproduce the second r-process peak (see the entry ``decays and fission'' in Fig.~\ref{onlydecays} and Fig.~\ref{fragments_abla}~\&~\ref{hfb_fragments}). However, the final (``original'') distribution in the case of FRDM fits the second peak nicely, while for the HFB mass model the peak is shifted by several mass units. From Fig.~\ref{freezeandfinal} it becomes clear that (not the overall abundance shape, but) the peak positions are in both cases close to the average peak position at (n,$\gamma$)-($\gamma$,n) freeze-out. This seems to indicate that even in these final phases an r-process path is again established which is closer to stability for the HFB than for the FRDM mass model, leading to a peak shifted to higher masses (for the conditions obtained in the dynamical ejecta of neutron star mergers). This effect can only be avoided either by a change of the nuclear mass model or by different environment conditions.

\begin{figure*}[t!]
  \hspace{-1cm}
  \includegraphics[width=0.55\textwidth]{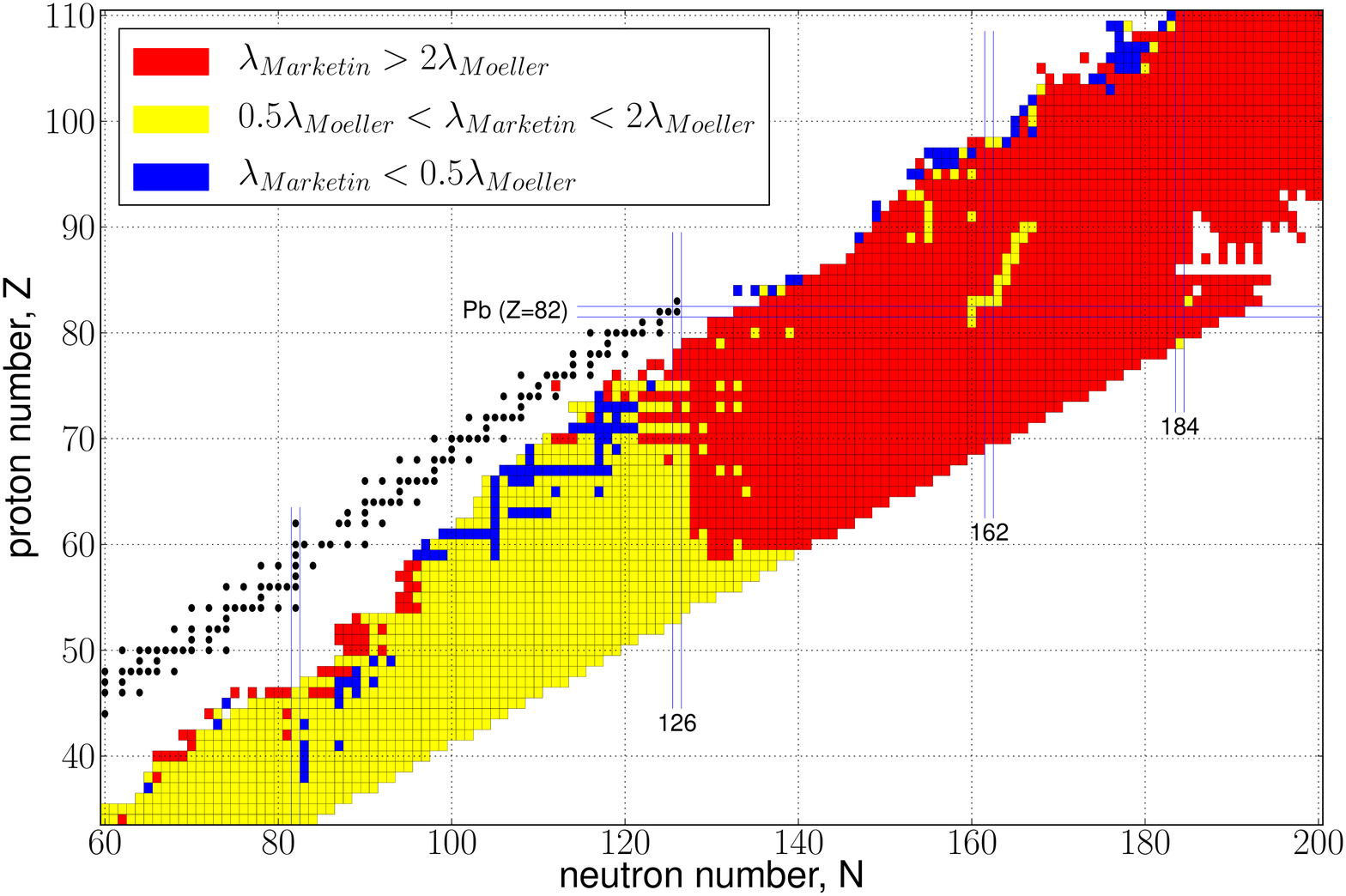}
  \label{marketin_betarates_qual}
  \includegraphics[width=0.55\textwidth]{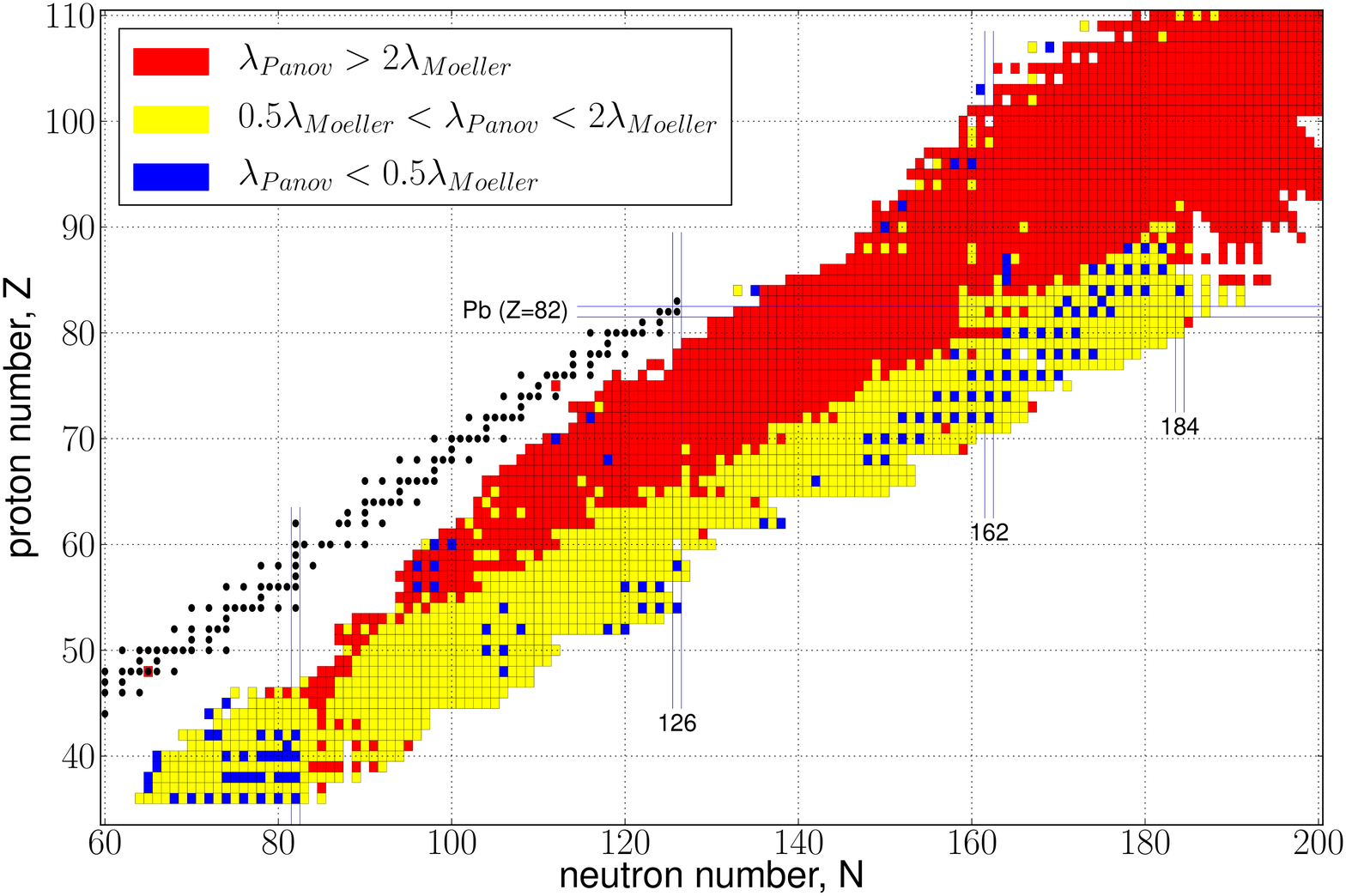}
  \label{panov_betarates_qual}
  \caption{Left panel: Comparison of the new \cite{marketin2015} $\beta$-decay rates with the old \cite{moeller2003} rates. A red square means that the \cite{marketin2015} $\beta$-decay rate ($\lambda_{Marketin}$) of the corresponding nucleus is more than two times faster than the \cite{moeller2003} rate, while a blue square signifies that the \cite{marketin2015} rate is slower than the \cite{moeller2003} rate by more than a factor of 2. If the two rates are within a factor of 2 to each other, the square is coloured yellow. Right panel: Same for the new \cite{panov2015} rates.}
\label{betarates_comparison}
\end{figure*}

\begin{figure*}[p!]
%  \centering
  \includegraphics[width=0.55\textwidth]{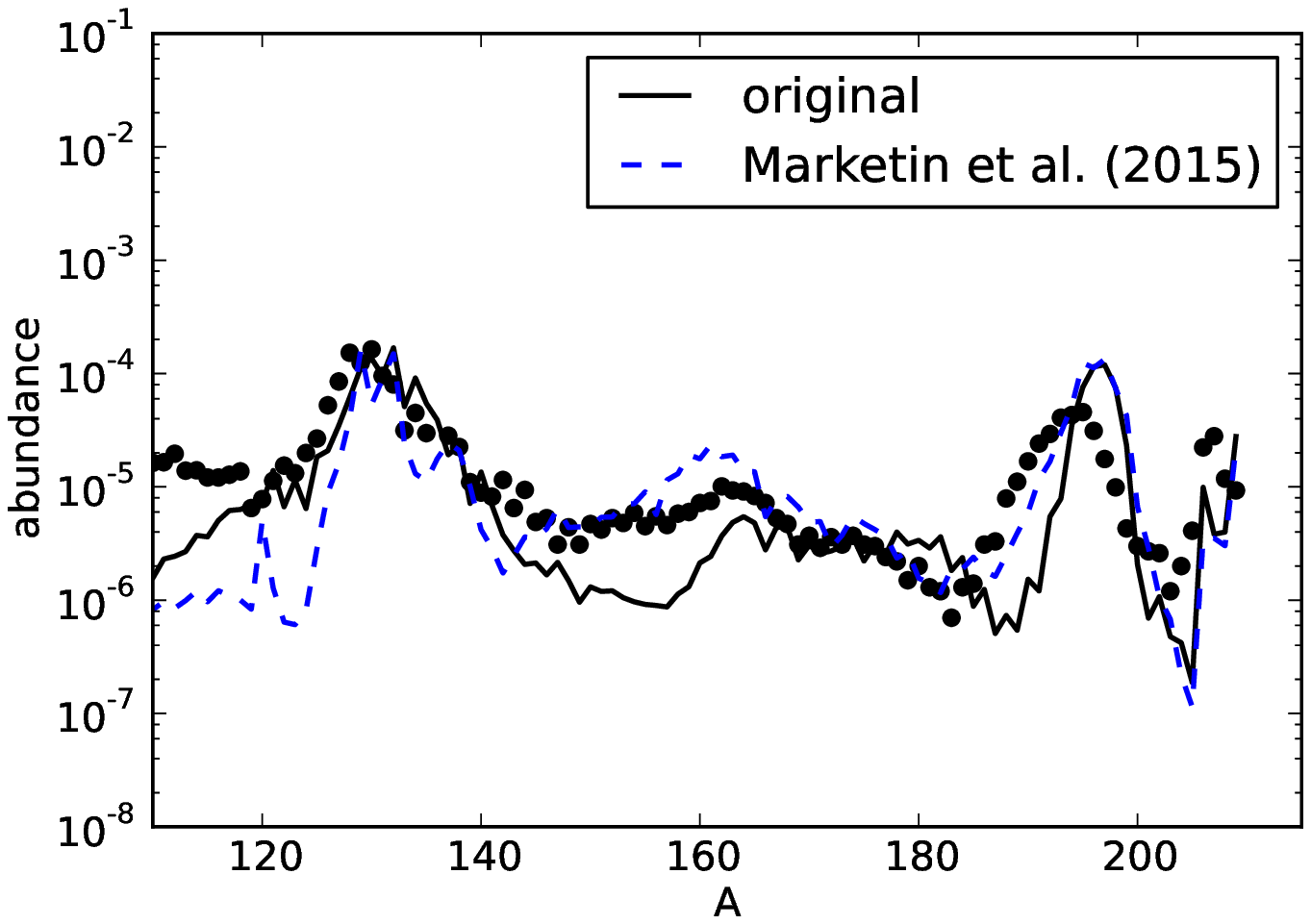}
  \label{marketin_frdm}
  \includegraphics[width=0.55\textwidth]{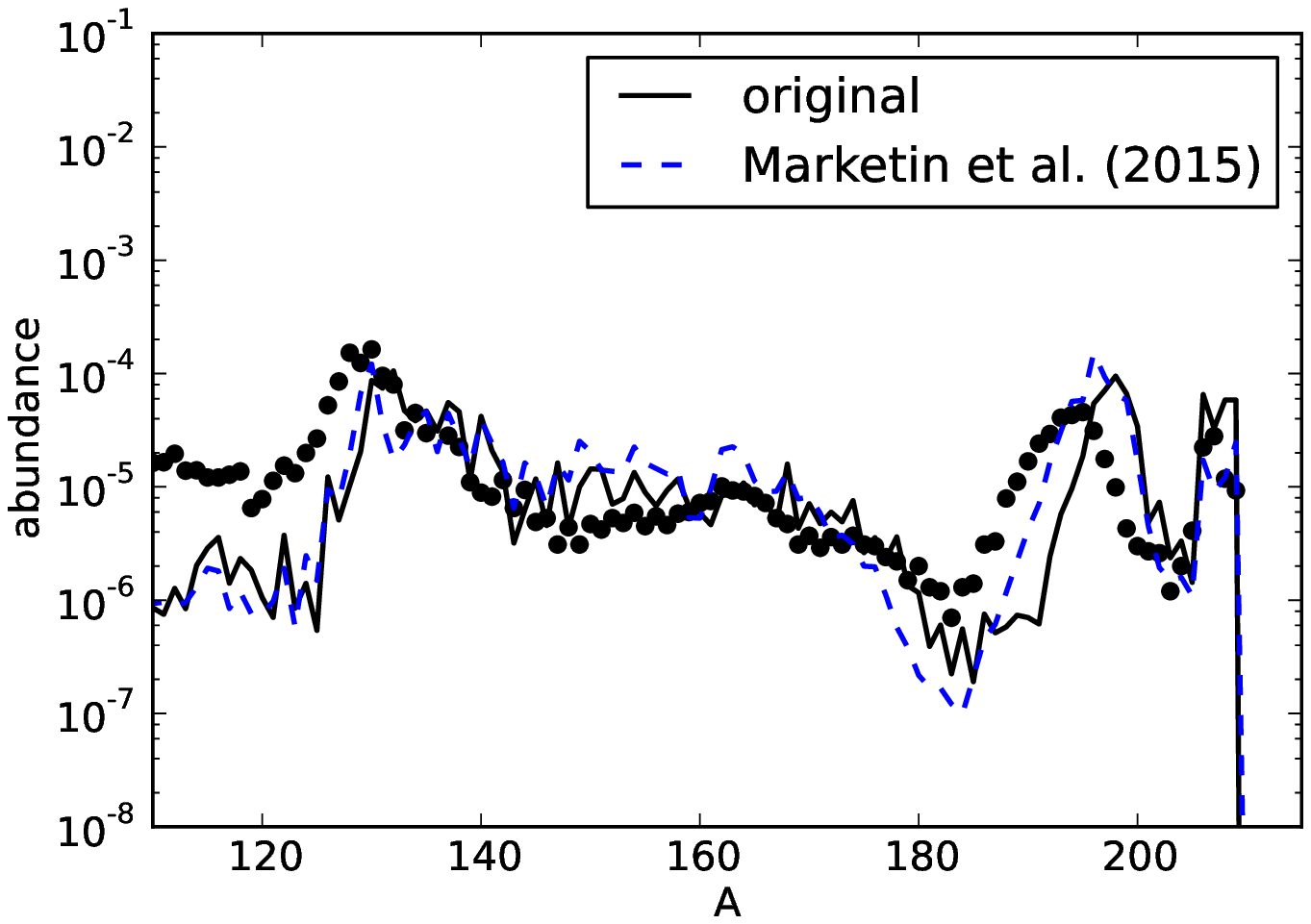}\\
  \label{marketin_hfb}
  \includegraphics[width=0.55\textwidth]{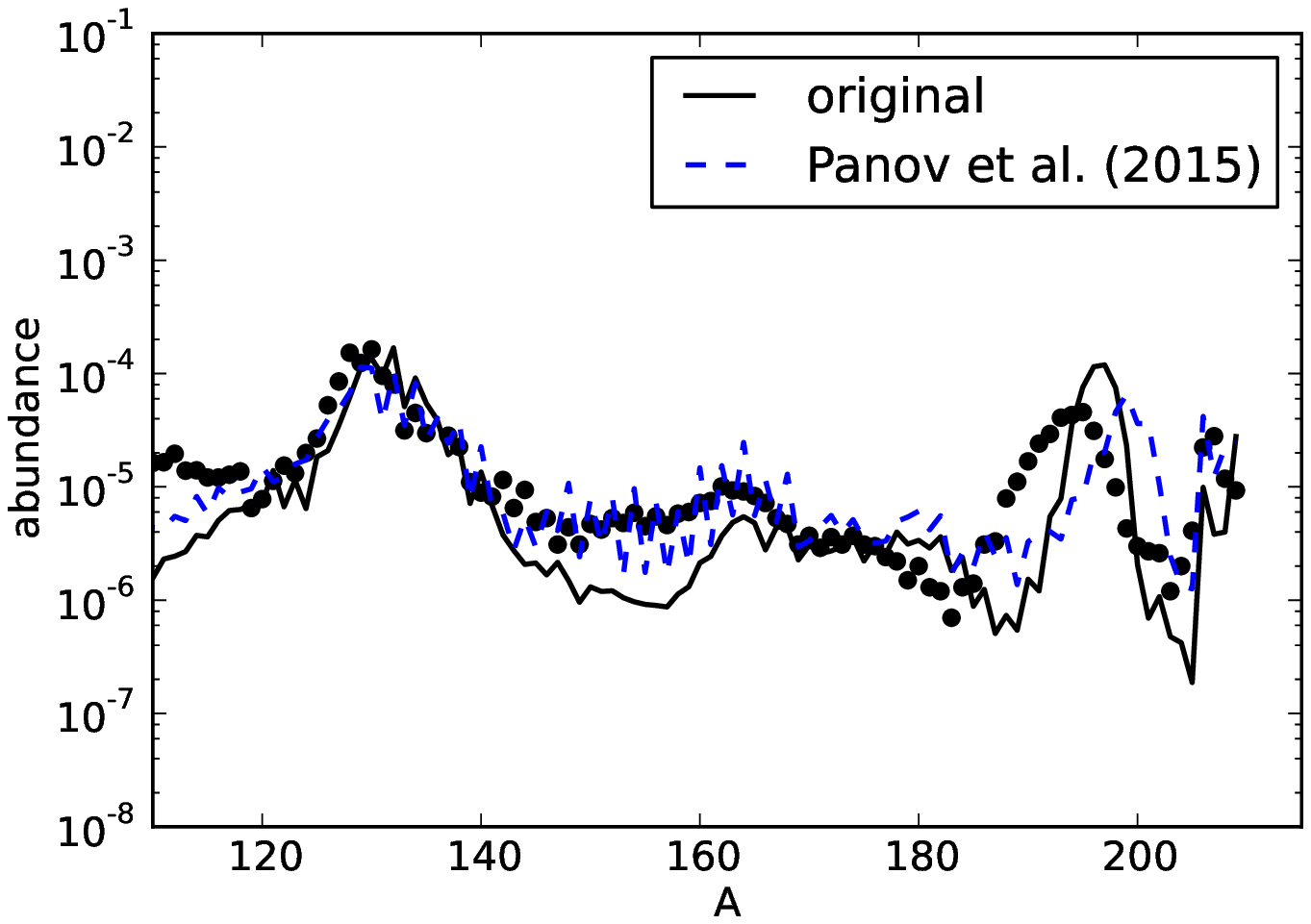}
  \label{panov_frdm}
  \caption{Top left: Final abundance distribution for a calculation using the \cite{marketin2015} rates together with the FRDM mass model and ABLA07. As a reference the FRDM, ABLA07 calculation from Fig.~\ref{ffdm24} is included. Top right: Same, but using the HFB-14 mass model. Bottom panel: Same as top left, but with the \cite{panov2015} rates.}
\label{marketin_panov_rates}
\end{figure*}

\subsection{Testing the Global Fit via Variations in Mass Models and Beta-Decay Rates}
\label{global}
Having shown the impact of a simple (and artificial) change in $\beta$-decay half-lives in the previous chapter, we now employ the newly calculated sets of half-lives of \cite{panov2015} and \cite{marketin2015}. Both new sets predict shorter half-lives for the majority of neutron-rich nuclei in the nuclear chart compared to the previously used \cite{moeller2003} half-lives. However, there are some decisive differences. In Figure~\ref{betarates_comparison} we present a comparison of the new $\beta$-decay rates with the \cite{moeller2003} rates that we have used before. The \cite{panov2015} set does not predict significantly faster rates far from stability, but in fact even noticeably slower rates (marked in blue) around $N=162$ close to the neutron drip line. The faster rates (red) closer to stability only come into effect after freeze-out. The \cite{marketin2015} calculations, on the other hand, predict faster rates for all nuclei on the r-process path beyond $N=126$. The impact on the final abundances can be seen in Figure~\ref{marketin_panov_rates}, where we present calculations performed using the \cite{marketin2015} rates combined with both the FRDM and HFB-14 reaction sets as well as the \cite{panov2015} rates together with the FRDM model. Note that the \cite{marketin2015} rates have been calculated using a different mass model, so they are not fully consistent with neither FRDM nor HFB-14. The \cite{panov2015} rates are based on FRDM, therefore we do not show a calculation with HFB-14.

The \cite{marketin2015} rates show a similar effect on the final abundances as our artificial study in Fig.~\ref{fastbeta}, broadening the low-mass flank of the third peak and increasing the abundances around the rare-earth peak. In fact, the broadening of the peak to lower mass numbers strongly improves the shape of the peak and (at least for the HFB-14 mass model) even a shift of the position to lower masses can be observed.
The \cite{panov2015} rates have a different effect. Here the $\beta$-decays are faster for nuclei with $N=126$ along the r-process path (before the freeze-out). Therefore the reaction flux proceeds faster in this region before it is held up afterwards at higher mass numbers, which means that less matter is accumulated in the peak. As a result, the height and shape of the third peak matches the solar peak very well (Fig.~\ref{panov_frdm}). However, as the abundances of the nuclei in the peak are lower by roughly a factor of 2, each nucleus in the third peak can capture double the amount of neutrons and the effect of the third peak shift is increased. Furthermore, the \cite{panov2015} rates show strong odd-even dependencies in the mass region $140 < A < 170$ (Fig.~\ref{panov_betarates_qual}), a quality which is reflected in the final abundances in this mass region.

\section{Conclusions and Outlook}

In the present paper we have tested (for a given set of astrophysical conditions related to the dynamic ejecta of neutron star mergers) the effects of the complex interplay of the nuclear input, including mass models, fission and fission fragment distributions as well as beta-decay half-lives.

We have shown that the r-process yields are strongly affected by fission and the adopted model for fission fragment distributions. In general, we find that more sophisticated fission fragment distribution models (ABLA07) improve the overall agreement with the solar r-process abundances. Similar studies with different fission fragment distribution models have been performed recently \citep{goriely2013,goriely2015}. Not surprisingly, the most significant variation is in the mass region $A = 100 - 160$, where the majority of the fission fragments is produced. This includes the second r-process peak and the rare-earth subpeak. Variations in nuclear mass models applied are decisive as well and we find that the combination of the applied mass model and the fission fragment distribution is essential for reproducing this mass region. In extreme cases of mass models which lead to fission only for $A > 300$ nuclei, the second r-process peak might not be produced at all \citep{shibagaki2015}.

In neutron-rich NSM nucleosynthesis, the third peak in the final abundance distribution shifts
towards higher masses, if after the (n,$\gamma$)-($\gamma$,n) freeze-out the conditions for further neutron captures of neutrons released during fission prevail. If the neutron density is still sufficiently high, several neutron captures after freeze-out can shift the peak. It is possible that for mass models not utilized in this study, which have the third peak shifted to lower masses in (n,$\gamma$)-($\gamma$,n) equilibrium (see e.g. \citealt{mendoza2014}), the final neutron captures shift the peak to its correct position.
 
We have also tested the effect of increased $\beta$-decay rates for the heaviest nuclei in our network ($Z \geq  80$), in which case the reaction flux is accelerated, leading to an earlier release of the fission (and $\beta$-delayed) neutrons which are recycled in the (n,$\gamma$)-($\gamma$,n) equilibrium that is present before the freeze-out, and leaving less matter in the fissioning region during freeze-out. This effect was further tested for new sets of theoretical $\beta$-decay
half-lives \citep{panov2015,marketin2015}, which can lead to very different results concerning the shape and position of the third peak. It should be noted that apparently the faster $\beta$-decay rates tested here can reduce the amount of late neutron captures and the shift of the third peak, but (at least in the present calculations) the shift of the third peak would not be prevented completely.

In summary, we explored the complex interplay of mass models, fission, and beta-decay half-lives for a variety of nuclear iputs and their impact on the resulting overall r-process abundances. Further changes with new versions of the FRDM model \citep{moeller2012,kratz2014} after their public release and many improved future nuclear structure predictions will be needed to settle these aspects completely.

Independent of the nuclear aspects/uncertainties studied here, it should also be realized that the astrophysical conditions matter as well. Here we utilized only the conditions for the dynamical ejecta of neutron star mergers within the treatment of \cite{korobkin2012} or \cite{bauswein2013}. However, these dynamical ejecta can also be affected by neutrino interactions and NSM ejecta include, apart from the dynamical channel, also matter ejected via neutrino-driven winds (e.g., \citealt{dessart2009,rosswog2014,perego2014,just2014}) and matter from unbinding a substantial fraction of the late-time accretion disk (e.g., \citealt{metzger2008,beloborodov2008,lee2009,fernandez2013b,fernandez2013a,just2014}). These additional channels yield larger electron fractions, since matter stays substantially longer near the hot central remnant and therefore positron captures and neutrino absorptions are likely. A number of recent studies \citep{just2014,perego2014,wanajo2014} find a broader range of $Y_e$-values that may be beneficial for the production of r-process elements and may also contain
substantial ``weak'' r-process contributions (for a parametric study of possible neutrino and antineutrino luminosities and average energies see \citealt{goriely2015b}). These would be closer to conditions from investigations for matter ejected in the jets of magneto-rotationally powered core-collapse supernovae \citep{winteler2012}, leading to less fission cycling and less final neutron captures from fission neutrons. Both aspects, improvements in the nuclear structure input as well as the complete description of the astrophysical conditions encountered in NS-NS and also NS-BH mergers should be followed in the future.

This work was partially supported by the Swiss NSF, the European Research Council grant FISH, the Nuclear Astrophysics Virtual Institute (NAVI), the SCOPES programme for co-operation between Eastern Europe and Switzerland, the THEXO collaboration within the 7$^{th}$ Framework Programme of the EU, and the COST Action New Compstar. A. A. was supported by the Helmholtz-University Young Investigator grant No. VH-NG-825. O.K. and S.R. were supported by DFG grant RO-3399, AOBJ-584282 and by the Swedish Research Council (VR) grant 621-2012-4870. I.P. was furthermore supported in part by RFBR grant 13-02-12106-ofi\_m.

\addcontentsline{toc}{section}{\hspace{0.9cm}Bibliography}
\bibliographystyle{apj}
\bibliography{Masterbib_2}

\end{document}